\documentclass[letterpaper,twocolumn,10pt]{article}
\usepackage{usenix2019_v3}
\usepackage[width=7in, height=9in, left=0.75in, right=0.75in, top=1in, bottom=1in]{geometry}

\usepackage{tikz}
\usepackage{amsmath}
\usepackage[normalem]{ulem}
\usepackage{pifont}
\usepackage{makecell}
\usepackage{titlesec}
\usepackage{hyperref}
\usepackage{graphicx}
\usepackage{caption}
\usepackage{epstopdf}
\usepackage{svg}
\usepackage{adjustbox}
\usepackage{amsfonts}
\usepackage{array}
\usepackage{booktabs}
\usepackage{enumitem}
\usepackage[most]{tcolorbox}
\usepackage{threeparttable}    
\tcbset{fontupper=\small,width=0.9\linewidth}
\usepackage{xcolor}
\usepackage{multirow}
\usepackage{diagbox}
\usepackage{bm}
\usepackage{threeparttable}
\usepackage{subcaption}

\titleformat{\paragraph}[runin]{\normalfont\normalsize\bfseries}{\theparagraph}{1em}{}[]
\usepackage[linesnumbered,ruled,vlined]{algorithm2e}
\titleformat{\section}{\large\bfseries}{\thesection}{1em}{}

\newenvironment{packeditemize}{
\begin{list}{$\bullet$}{
\setlength{\labelwidth}{8pt}
\setlength{\itemsep}{0pt}
\setlength{\leftmargin}{\labelwidth}
\addtolength{\leftmargin}{\labelsep}
\setlength{\parindent}{0pt}
\setlength{\listparindent}{\parindent}
\setlength{\parsep}{0pt}
\setlength{\topsep}{3pt}}}{\end{list}}

\newcommand{\name}{\texttt{RedNet}}

\newcommand{\paragraphb}[1]{\vspace{0mm}\noindent\textbf{#1}}

\begin{document}
\date{}

\title{A Case for Application-Aware Space Radiation Tolerance in Orbital Computing
}

 \author{
 \normalsize {\rm Meiqi Wang$^{1}$, Han Qiu$^{1*}$, Longnv Xu$^{1}$, Di Wang$^{2}$, Yuanjie Li$^{1}$, Tianwei Zhang$^{3}$, Jun Liu$^{1}$, Hewu Li$^{1}$}\\
\normalsize {$^{1}$Tsinghua University, $^{2}$Beijing University of Posts and Telecommunications, $^{3}$Nanyang Technological University.}\\
\normalsize {\rm $^{*}$qiuhan@tsinghua.edu.cn}
 } 

\maketitle

\begin{abstract}
We are witnessing a surge in the use of commercial off-the-shelf (COTS) hardware for cost-effective in-orbit computing, such as deep neural network (DNN) based on-satellite sensor data processing, Earth object detection, and task decision.
However, once exposed to harsh space environments, COTS hardware is vulnerable to cosmic radiation and suffers from exhaustive single-event upsets (SEUs) and multi-unit upsets (MCUs), both threatening the functionality and correctness of in-orbit computing.
Existing hardware and system software protections against radiation are {\em expensive} for resource-constrained COTS nanosatellites and {\em overwhelming} for upper-layer applications due to their requirement for heavy resource redundancy and frequent reboots.
Instead, we make a case for cost-effective space radiation tolerance using application domain knowledge. 
Our solution for the on-satellite DNN tasks, \name, exploits the uneven SEU/MCU sensitivity across DNN layers and MCUs' spatial correlation for lightweight radiation-tolerant in-orbit AI computing.
Our extensive experiments using Chaohu-1 SAR satellite payloads and a hardware-in-the-loop, real data-driven space radiation emulator validate that \name~can suppress the influence of radiation errors to $\approx$ 0 and accelerate the on-satellite DNN inference speed by 8.4\%--33.0\% at negligible extra costs.
\end{abstract}

\section{Introduction}
\label{sec:intro}

Low Earth Orbits (LEOs) are evolving into a new class of ``mobile computing clusters''.
The recent technological advances in rocket reusability and affordable nanosatellites have sharply decreased the cost of launching computing payloads to space, stimulating a surge in operational satellite constellations equipped with cameras, sensors, and GPUs~\cite{esposito2019highly,adams2019towards,starDetect}. 
These in-orbit edge computing resources empower satellites to locally process runtime raw data (e.g., 20TB/day of Earth observation images, sensor data, and system logs) rather than transmitting them to the ground stations for remote processing, thus saving precious satellite downlink bandwidth \cite{cappaert2018building,denby2020orbital,denby2019orbital,denby2023kodan,tao2024known} and facilitating real-time satellite tasks like disaster detection~\cite{leyva2023satellite}, climate monitoring~\cite{wiseman2019impact}, and precision agriculture~\cite{kong2019monitoring}. 

A key factor for in-orbit computing's commercial success lies in the capital cost of satellites.
To this end, recent efforts from industry (e.g., ESA's Phi-Sat-1~\cite{esposito2019highly} and NVIDIA Jetson Xavier NX GPU in space~\cite{pri2024}) and academia \cite{denby2020orbital,denby2023kodan,wang2023mars,xing2024deciphering} advocate building satellites using commercial off-the-shelf (COTS) hardware rather than dedicated special-purpose space hardware, which is not only 10$\times$ cheaper \cite{denby2020orbital,denby2019orbital,xing2024deciphering} 
but also more powerful to support computing-intensive tasks like deep neural network (DNN)-based sensory data processing, Earth object detection, and satellite control decision~\cite{esposito2019highly,denby2023kodan,tao2024known, choumos2024artificial}.

However, COTS hardware-powered in-orbit computing is prone to errors when exposed to harsh space environments.
Unlike the terrestrial environment, {\em space radiation} is a norm in LEOs rather than an exception.
Without radiation hardening in expensive dedicated space hardware, COTS electronic components will suffer from exhaustive disturbances by such radiation.
Both in-orbit experimental records~\cite{bezerra2011carmen2,samaras2011carmen} and ground radiation tests~\cite{walters2013practical,fabero2020single} have confirmed that space radiation particles can flip COTS memory bits (i.e., 0 $\rightarrow$ 1 or 1 $\rightarrow$ 0) at least {\em hundreds of times per day}.
Such frequent bit flips are fatal for the basic functionality and correctness of in-orbit computing, especially for memory-intensive DNN tasks in satellites whose only one-bit flip in neural network parameters can lead to significant inference errors.

To combat these space radiation-induced errors, state-of-the-art solutions have explored general-purpose protections based on
physical hardware redundancy~\cite{su2021triple} (e.g., 3 duplicate flight computers in each SpaceX rocket for majority vote~\cite{spaceXRocket}),
error correction code~\cite{zhou2022carbink,wang2023mars, schmidt2017radiation, skeggs2022vivid}, or software reboot~\cite{saramud2022implementation}. 
But from the in-orbit task perspective, these general-purpose hardware and system software protections are {\bf expensive} and {\bf overwhelming} for resource-constrained COTS nanosatellites.
On the one hand, to tolerate frequent bit errors, extra redundancies must be used in the hardware (thus increasing satellite cost) or software (whose error correction code redundancy can even exceed the original data length).
On the other hand, correcting all radiation-induced errors can be {\em more than necessary} for many tasks.
As we showcase with DNN-based tasks in \S\ref{sec:dnnanalysis}, the underlying bit flip-induced DNN parameter error may not always affect DNN task results.
Correcting these errors at high costs does not yield benefits for applications.

In this work, we explore an alternative {\em application-aware} space radiation tolerance paradigm.
Our key finding is that application-aware radiation tolerance can be more effective and affordable for COTS-based satellites for two reasons (\S\ref{sec:dnnanalysis}).

\begin{packeditemize}
\item {\bf Not all bits are equal:} In the application view, some bit errors (e.g., in a DNN model's shallow layers) are much more critical than others (e.g., in a DNN's deep layers).
By distinguishing these bits and handling them differently with domain knowledge, applications can be protected with fewer hardware/software redundancies at lower costs. 

\item {\bf Spatial correlation across multi-bit errors:}
A salient feature of radiation-induced bit errors is their simultaneous occurrence, incurring frequent multi-bit errors in addition to single-bit errors.
To tackle these multi-bit errors, existing general-purpose hardware/software protections must pay more redundancies/time.
Instead, we note that those multi-bit errors exhibit strong spatial contiguity, thus tolerable at lower costs with application-specific knowledge. 

\end{packeditemize}

To this end, we make a case for application-aware space radiation tolerance using DNN-based tasks, one of the most important applications in orbital computing.
Our solution, \underline{R}adiation \underline{E}rror-tolerant \underline{D}eep neural \underline{Net}work (\name),
follows the above insights to rearrange the DNN model for low-cost radiation tolerance (\S\ref{sec:design}).
\name~renovates the classical activation function to suppress inference error propagation from DNN's shallow layer to the deep layer.
\name~also adopts a multi-exit strategy to allow early exit at inference to mitigate multi-bit errors in sensitive DNN layers.

We experiment \name~with the NVIDIA TensorRT and the NVIDIA Jetson Xavier NX (deployed as a payload on Chaohu-1 SAR satellite~\cite{Chaohu-1} launched in 2022). 
We use this payload (see \figurename~\ref{fig:crash} in \S\ref{sec:eval}) manufactured by~\cite{starDetect} for testing. 
Moreover, we develop a hardware-in-the-loop, end-to-end space radiation error emulator to conduct extensive experiments with 3 satellite image datasets and 3 DNN model structures. 
Both experimental results validate that \name~can not only suppress the influence of radiation-caused errors close to 0 but also can speed up the inference by 8.4\%-33.0\% with a tiny extra memory footprint and less satellite energy. 

\smallskip
\paragraphb{Artifact release.} We open-source our end-to-end radiation emulator\footnote{\url{anonymous.4open.science/r/radiation-error-emulator-5F89}} and \name~artifacts\footnote{\url{anonymous.4open.science/r/RedNet-9665}}.

\section{Motivation}
\label{sec:motivation}

\subsection{Space Radiation Hazards for Satellites}
\label{sec:motivation:hazards}

\begin{figure}[t]
\subfloat[Characterization and modeling of the environment missions.\label{subfig:radiation_test_a}]{
    \begin{minipage}[t]{\linewidth}
        \centering
        \includegraphics[width=0.99\linewidth]{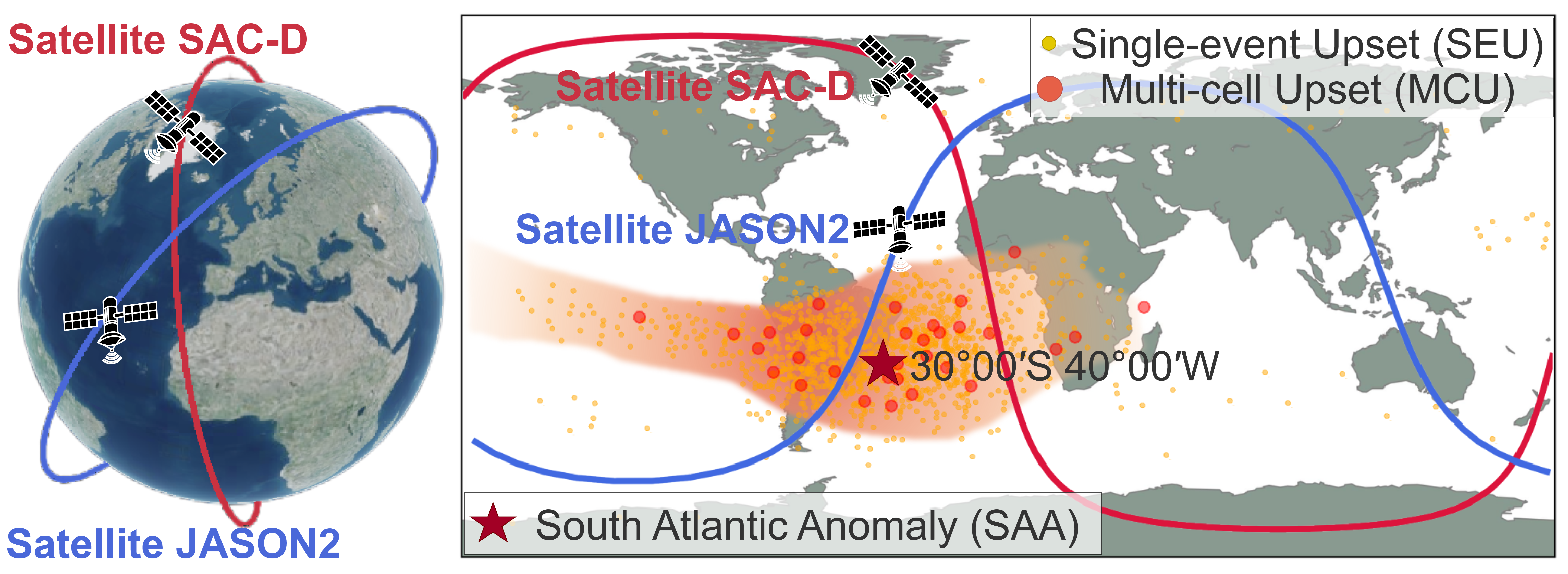}
    \end{minipage}
    }
    \\
    \subfloat[Error occurrences in JASON2.\label{subfig:radiation_test_b}]{
    \begin{minipage}{0.48\linewidth}
        \centering
        \includegraphics[width=\linewidth]{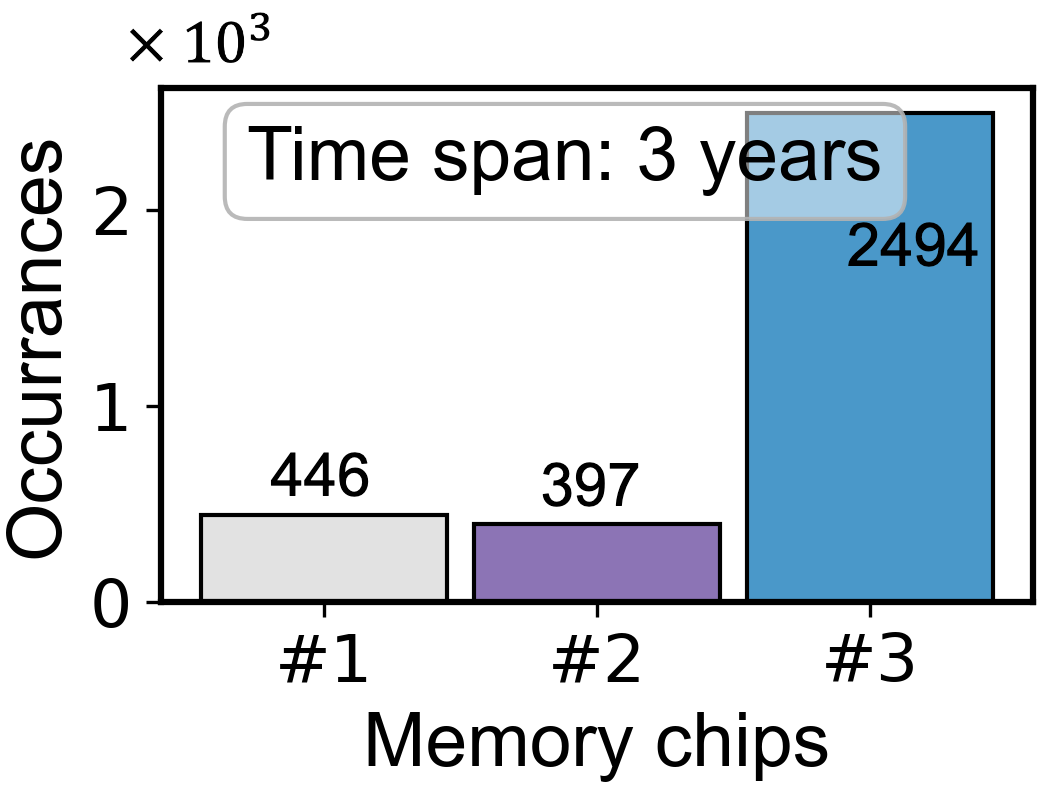}
    \end{minipage}
    }
    \subfloat[Missions in LEO.\label{subfig:radiation_test_c}]{
    \begin{minipage}{0.49\linewidth}
        \centering
        \includegraphics[width=\linewidth]{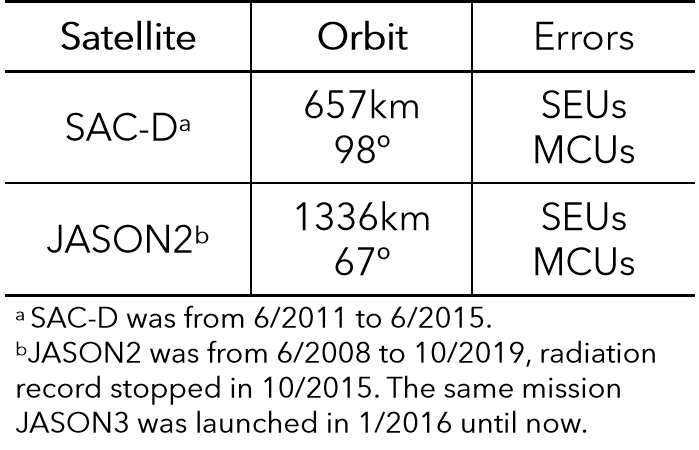}
    \end{minipage}%
    }
    \vspace{-1ex}
    \caption{Summary of existing radiation tests in space~\cite{bezerra2011carmen2}.}
    \label{fig:radiation_test}
    \vspace{-2ex}
\end{figure}

{Harsh space radiation} has emerged as a potent ``adversary'' for in-orbit satellite applications.
Satellites in outer space are surrounded by cosmic rays, solar radiation, and high-energy charged particles that will cause \textit{permanent damage} (e.g., degradation and burning) or \textit{transient damage} (e.g., logical state flip in memory) in electronic components on satellite payloads, especially the COTS devices~\cite{wang2023mars}.
Permanent damages, such as single-event latch-up and total ionizing dose, are relatively easy to tackle because they can be detected through hardware- or system-extractable features~\cite{yue2017single} and subsequently restored to prevent further deterioration. 

Instead, {\em transient damages}, although seemingly trivial, are more prevalent and threatening for COTS-based in-orbit computing.
They can temporarily flip logical states in the memory~\cite{caron2019physical,nidhin2017understanding},
inject wrong control logic and data values into the application process,
and disrupt the in-orbit computing outcomes.
Although non-destructive for hardware, these radiation-caused memory errors are more challenging to deal with since they can be {\em stealthy}:
These errors could go {\em undetected} by hardware and system until the application ``sees'' (as we will detail and empirically validate in \S~\ref{subsec:senanaly}). 
It is difficult to detect these silent errors effectively in applications.  

\begin{figure}[t]
\subfloat[Illustration of bit-flip errors in satellite.\label{subfig:seumcu_a}]{
    \begin{minipage}{0.71\linewidth}
        \centering
        \includegraphics[width=0.98\linewidth]{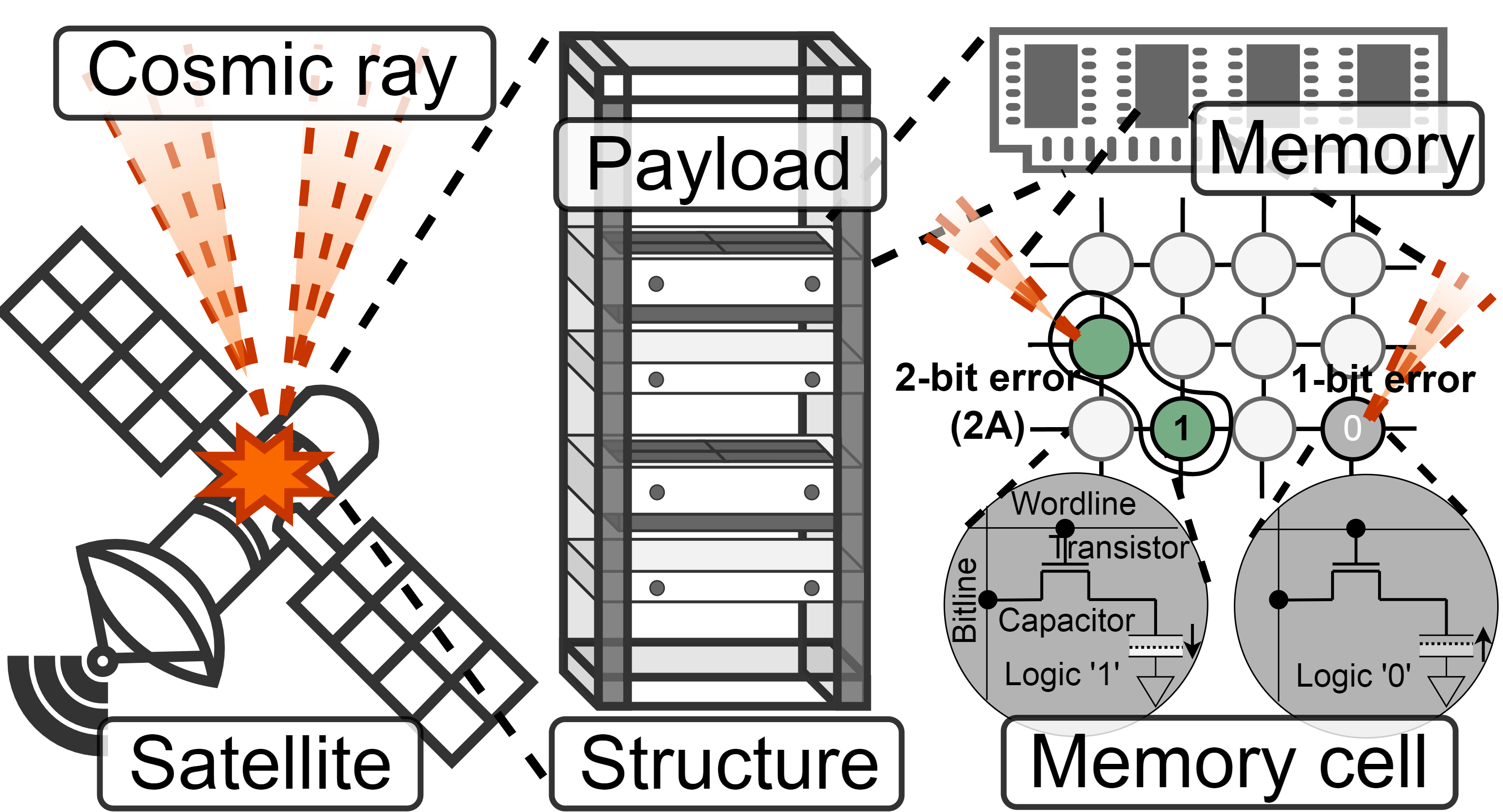}
    \end{minipage}
    }
    \hspace{-2mm}
    \subfloat[2-bit error.\label{subfig:seumcu_b}]{
    \begin{minipage}{0.26\linewidth}
        \centering
        \includegraphics[width=0.96\linewidth]{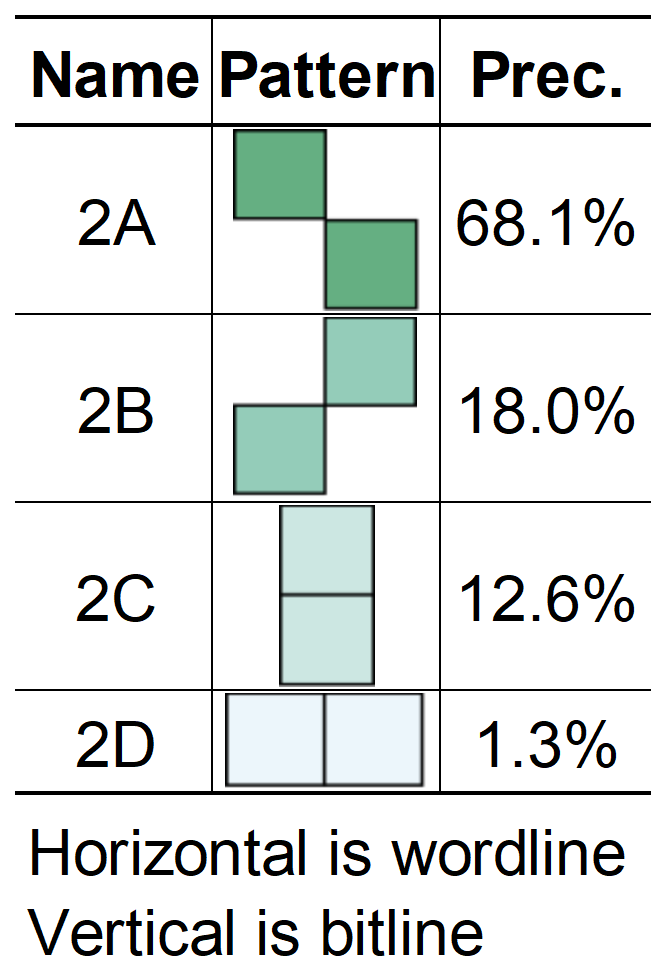}
    \end{minipage}
    }
    \\
    \subfloat[Occurrences and distribution of different error patterns.\label{subfig:seumcu_c}]{
    \begin{minipage}{\linewidth}
        \centering
        \includegraphics[width=\linewidth]{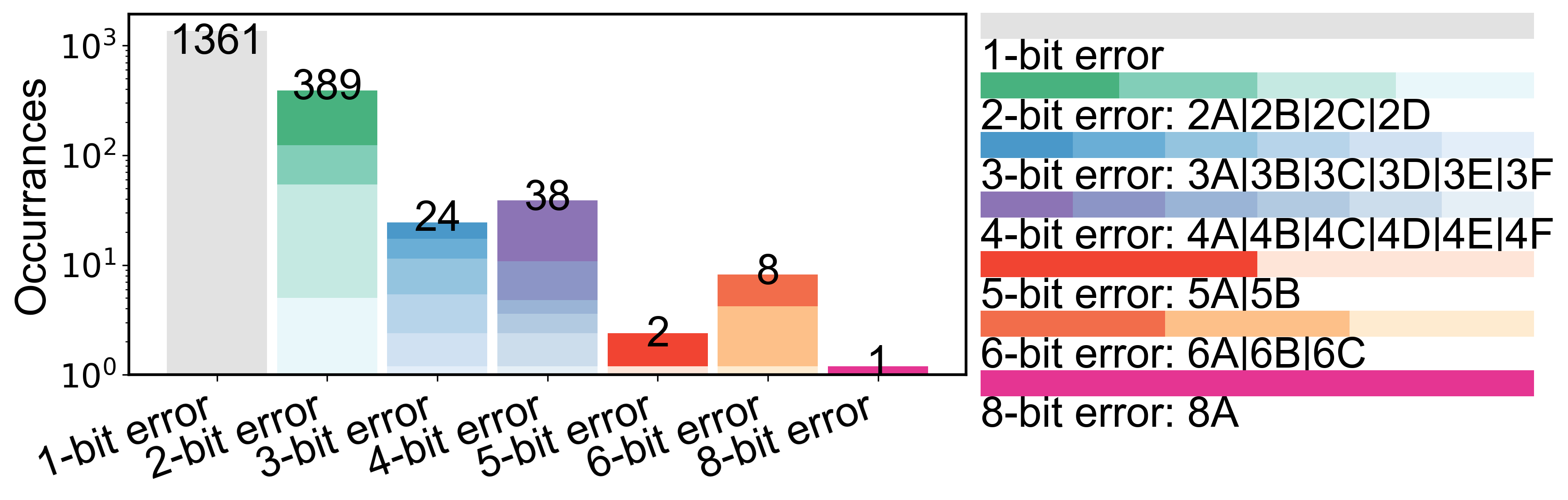}
    \end{minipage}
    }
    \vspace{-1ex}
    \caption{Occurrences and patterns of bit-flip errors~\cite{fabero2020single}.}
    \label{fig:seumcu}
    \vspace{-2ex}
\end{figure}

In operational satellites, space radiation-induced transient damages have been frequently perceived by satellites. 
\cite{bezerra2011carmen2, samaras2011carmen, bezerra202314} deploys instruments on satellites to record in-orbit radiation-caused memory bit errors for different COTS memory chips. 
\figurename~\ref{fig:radiation_test} illustrates multi-memory-chip bit error records in two different LEO satellites (i.e., satellite SAC-D\footnote{\url{https://podaac.jpl.nasa.gov/aquarius}} and satellite JASON2\footnote{\url{https://www.jpl.nasa.gov/missions/jason-2}} in \figurename~\ref{subfig:radiation_test_c}) and \figurename~\ref{subfig:radiation_test_a} shows an approximate location where those bit errors occurred.
The South Atlantic Anomaly (SAA) is the most severely affected region and the radiation effect in orbits with different altitudes and inclinations is not always the same (e.g., the SAA at 657 km is less extended than it is at 1336 km~\cite{bezerra202314}). 
\figurename~\ref{subfig:radiation_test_b} shows a record of 3-year radiation-caused memory bit error occurrences of different memory chips
recorded in~\cite{samaras2011carmen} on JASON2, which are about
$4.76\times10^{-7}$/bit/day. 
This indicates that for one typical application that occupies a 40MB memory footprint, there will be roughly 150+ bit errors happening per day.
Moreover,~\cite{dutta2007multiple} points out that more than 5\% radiation-caused bit errors can cause spatial-correlated multi-bit errors. 
This is due to multiple adjacent memory cells being influenced together leading to various special error patterns (details in \S~\ref{subsec:radiationThreats}).
Thus, high-frequent memory bit errors in space make reliable orbital computing more critical and urgent.

\subsection{Radiation Threats for Orbital Computing}
\label{subsec:radiationThreats}

Although radiation-caused bit errors in memory have occasionally been discovered on the ground, their rarity does not pose a major concern~\cite{zaharia2012resilient}. 
Instead, the frequent bit errors on COTS devices in space do cause significant challenges in deploying applications in orbit. 
In reality, radiation-caused bit errors can be classified into two categories:

\noindent{{\bf Single-bit errors (SEUs)\footnote{Single-Event Upsets (SEUs) \cite{yue2017single} in the aerospace jargon.}}} are induced by ionizing radiation striking that makes electron-hole pairs in transistor gates generate and diffuse. Single-bit errors are naturally present in the hazardous space environment and may alter the operation of memory.
\cite{underwood1990orbit,radaelli2005investigation, samaras2011carmen} studied the radiation effects in space and their influence on various space devices.
The probability of single-bit errors can reach $10^{-7}$ to $10^{-6}$ per bit per day~\cite{wang2023mars}.
Even a single-bit error may damage the performance of in-orbit applications (see detailed analysis in \S~\ref{subsec:unevensensitive}).

\noindent{{\bf Multi-bit errors (MCUs)\footnote{Multiple-Cell Upsets (MCUs) \cite{dutta2007multiple, kobayashi2020scaling} in the aerospace jargon.}}}
occur when neutron particles react with impurities and produce a large number of secondary particles with sufficient energy~\cite{dutta2007multiple, kobayashi2020scaling}, leading to multiple bit-flip errors simultaneously.
Multi-bit errors happen more frequently in the space COTS hardware due to the advanced nanotechnology and more compact chips in size~\cite{dutta2007multiple, aditya2018seu}.
Radiation can easily cause an arbitrary number of transient and silent bit errors in widely-used DRAM-based memory~\cite{kohler2017analysis, samaras2011carmen}.
For instance, \figurename~\ref{subfig:seumcu_c} summarizes the multi-bit errors (from 2 to 8) in radiation tests~\cite{fabero2020single}, 
reflecting the high frequency of multi-bit errors and the distribution of error patterns. 

We emphasize that {\bf multi-bit errors in space are {\em not} equal to multiple independent single-bit errors}.
Radiation can cause {\em spatial-correlated} multiple memory cells and flip the stored bits together. 
From the hardware/system perspective, a spatial-correlated multi-bit error seems similar to multiple single-bit errors simultaneously.
However, spatial correlation in memory must be considered since they have different influences on applications.
As we will reveal in Section~\ref{subsec:seuvsmcu}, different error patterns at the hardware layer 
will propagate to upper layers (see details in Appendix~\ref{appen:dram}) and cause diverse influences on applications.

\subsection{Why Not Classical Radiation Protections?}
\label{subsec:classicalProtections}

Radiation-induced errors have been a well-known issue for space computing.
Despite various radiation protections at the hardware and system software, they resurge recently with the use of COTS devices and DNN-powered applications in space.
We next discuss existing solutions and their limitations.

\noindent\textbf{{Hardware protection.}} 
The radiation hardening of space device~\cite{wang2023mars,geist2023nasa,leppinen2017current}  usually requires significantly reduced clock speed and increased die areas~\cite{burcin2002rad750} which leads to lagged CPUs, small memory, and outdated operating systems. 
For instance, EnduroSat OBC\footnote{\url{https://satsearch.co/products/endurosat-onboard-computer}}, a widely-used radiation-hardened computer, is powered by an ARM Cortex M4/M7 processor but only has 180/216MHz clock speed and support 2MB memory. 
This is far from supporting applications like DNNs~\cite{wang2023mars}. 
The other common way to mitigate radiation-induced bit errors in space is to introduce physical redundancy (e.g., triple modular redundancy (TMR)~\cite{su2021triple}) or error correction code memory~\cite{zhou2022carbink} at the hardware layer. 
For instance, SpaceX uses 3 processors for each of the 18 different processing units on their rockets (e.g., dragon cargo vessel~\cite{spaceXRocket}). 
However, the redundancy of 3 processors on rockets is not optimal to adopt on nanosatellites since it will introduce multiple times of energy consumption. 
The error correction code memory chip is not widely deployed in COTS devices since supporting it requires not only changing the memory chip but also the CPU and motherboard (e.g. Intel disables error correction code memory support in all memory controllers in desktop CPUs like I7). 
Moreover, using error correction code memory chips will not only introduce unignorable overhead~\cite{wang2023mars} but also cannot completely mitigate the radiation-caused bit errors~\cite{dutta2007multiple}.
 
\begin{figure}[t]
    \centering
    \includegraphics[width=\linewidth]{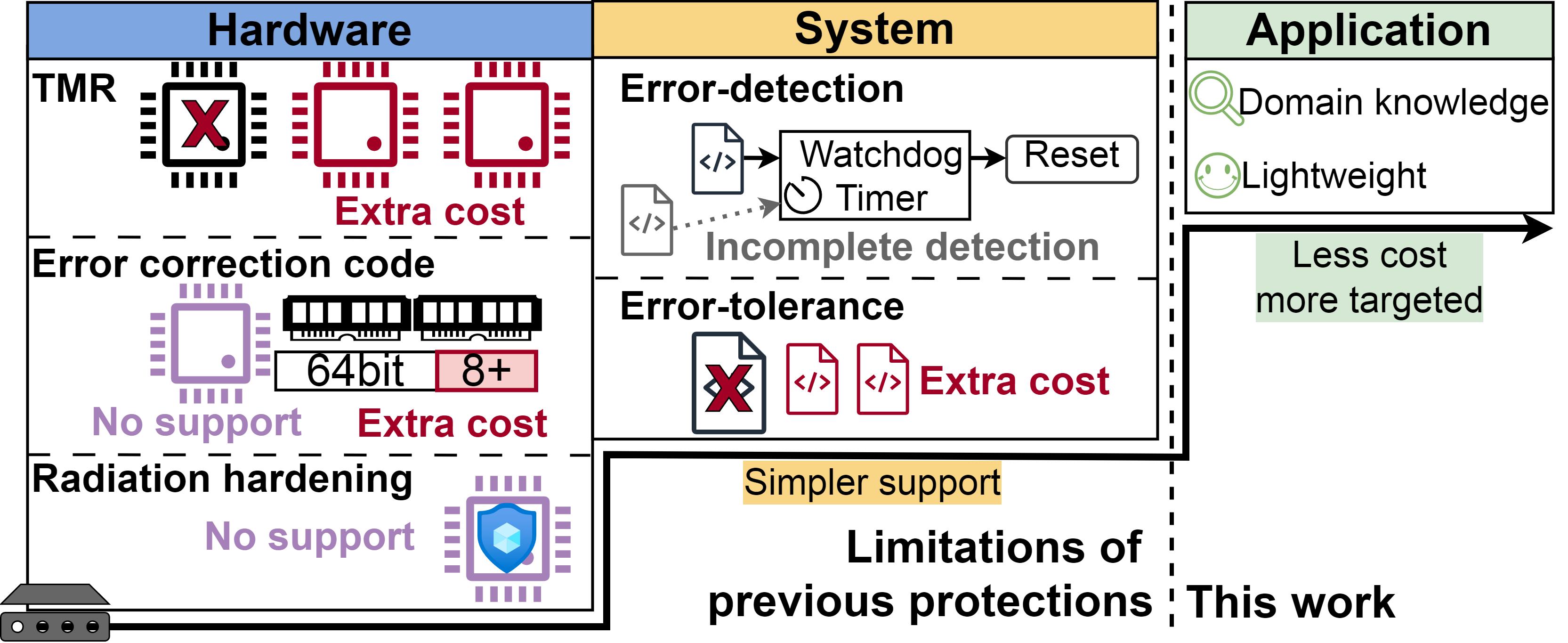}
    \caption{Limitations of classical space radiation protections.}
    \label{fig:protections}
    \vspace{-3ex}
\end{figure}

\noindent\textbf{System software protection} mainly includes error-detection and error-tolerance~\cite{dopson2005softecc, schmidt2017radiation, saramud2022implementation, skeggs2022vivid}.
For instance, a watchdog to detect bit errors will send a rebooting command and reload the data in memory when the application hangs or crashes~\cite{saramud2022implementation}.
However, bit errors in some applications, such as DNN models, usually do not cause hangs or crashes but only degrade the prediction accuracy~\cite{favalli2004annotated,wu2019protecting,saileshwar2022randomized}.      
For error tolerance, software error correction code~\cite{dopson2005softecc} is leveraged to repeatedly compute page-level checksums in the kernel, which is restricted to writing periods~\cite{dopson2005softecc}, and the correctable bits are limited by performance costs. 
For instance, correcting a 1-bit error in one byte (the most fine-grained tolerance) typically needs an extra 5-bit correction code, increasing 62.5\% memory usage~\cite{dopson2005softecc}. 
Besides, software-only redundancy includes the compiler generating two or more machine code sequences to run in multi-threads and compare outputs, which will inevitably multiply runtime and resource consumption~\cite{skeggs2022vivid}.

Generally, system software protection sacrifices performance to ensure global reliability for all upper-layer applications (i.e., bits of applications are equal to systems). 
However, treating upper-layer applications as black boxes is not always optimal, especially when bit errors are indiscriminately (over) protected (e.g., checking for \textit{every} bit error and making redundancy for the \textit{entire} application), resulting in unnecessarily high time and energy consumption, which is expensive for power-constrained nanosatellites.

\section{New Opportunities in the Application View}
\label{sec:dnnanalysis}

Motivated by the aforementioned deficiencies, 
we explore alternative lightweight space radiation protections with the assistance of application domain knowledge. 
We make such a case using one of the most important applications in orbital computing: DNN-based inference tasks (e.g., scene recognition and object detection in Earth observation satellites~\cite{esposito2019highly,tao2024known}). 

\begin{figure}[!htbp]
    \centering
    \includegraphics[width=\linewidth]{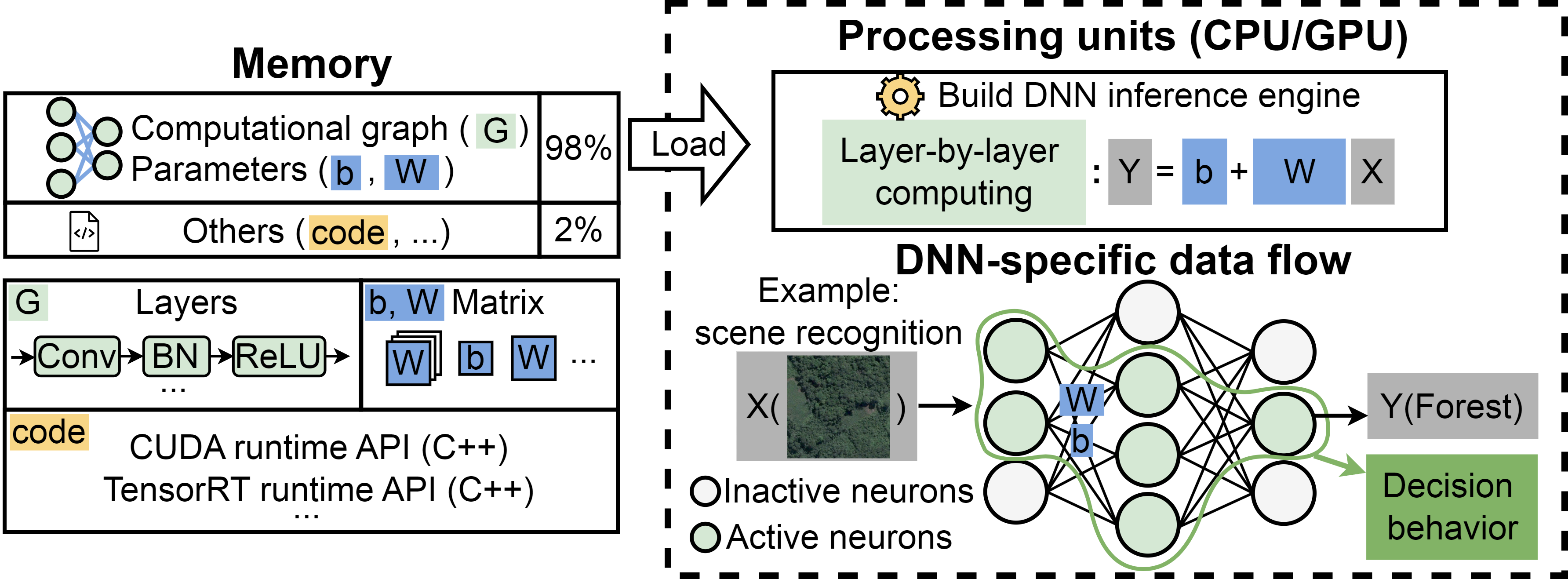}
    \vspace{-1ex}
    \caption{The runtime of DNN-specific applications.}
    \label{fig:DNNdataflow}
    \vspace{-3ex}
\end{figure}

\noindent\textbf{How DNN inference works in satellites:} 
{DNN has been broadly used in various in-orbit satellite computing tasks, such as humanitarian assistance, disaster detection~\cite{gupta2019xbd, leyva2023satellite}, climate monitoring~\cite{wiseman2019impact}, precision agriculture~\cite{kong2019monitoring}, and many more.}
{For each of these tasks, the corresponding} DNN model is usually first trained offline (i.e., on the ground) and then deployed onboard according to the satellite's payload systems. 
We list the details of how our trained models are converted and deployed on our payload system in \S~\ref{subsubsec:scenario}.
Then, the onboard model's runtime inference (\figurename~\ref{fig:DNNdataflow}) depends on DNN-specific data flow, which is determined by two parts.  
(1) The \textit{computational graph} (G) determines how typical DNN layers are arranged in memory like convolution (Conv) layers, batch norm (BN) layers, activation (e.g., ReLU) layers, etc.
(2) The \textit{trained parameters}, such as convolutional weights ($W$) and bias ($b$), are to be used for computing with the input of each layer.
The above two parts are pre-stored in the memory and then loaded in process units (e.g., GPU) for inference. 
At runtime, the inputs and the trained parameters will perform calculations specified by the computational graph and activate neurons layer by layer until the final results.

Transient bit errors during the runtime DNN inference tasks are harmful to safety-sensitive tasks~\cite{favalli2004annotated,wu2019protecting,saileshwar2022randomized} but are very difficult to perceive. 
This is because the DNN inference mainly depends on the memory-intensive data flow mentioned above, but not on branch instructions in the control flow graph like other applications~\cite{wang2023mars}. 
Radiation-caused bit errors in DNN-specific data flow will lead to wrong model parameters during inference calculation, which will only lead to wrong results without behaviors like hang or crash. 
Thus, radiation-induced bit errors can silently affect the runtime DNN models. 

In this section, we empirically analyze representative DNN-based orbital computing tasks (\S\ref{subsec:setup}) to reveal {the benefits of application-aware radiation protections}.
We unveil two opportunities in this direction: uneven radiation error sensitivity (\S\ref{subsec:unevensensitive}) and spatially correlated multi-bit errors (\S\ref{subsec:seuvsmcu}).
These benefits motivate us to explore the potential of application-aware space radiation tolerance in the next section.

\subsection{Experimental Setup}
\label{subsec:setup}

We empirically study representative DNN-based orbital tasks to motivate {application-aware radiation protections}.  

\begin{figure}[!htbp]
    \centering
    \includegraphics[width=0.98\linewidth]{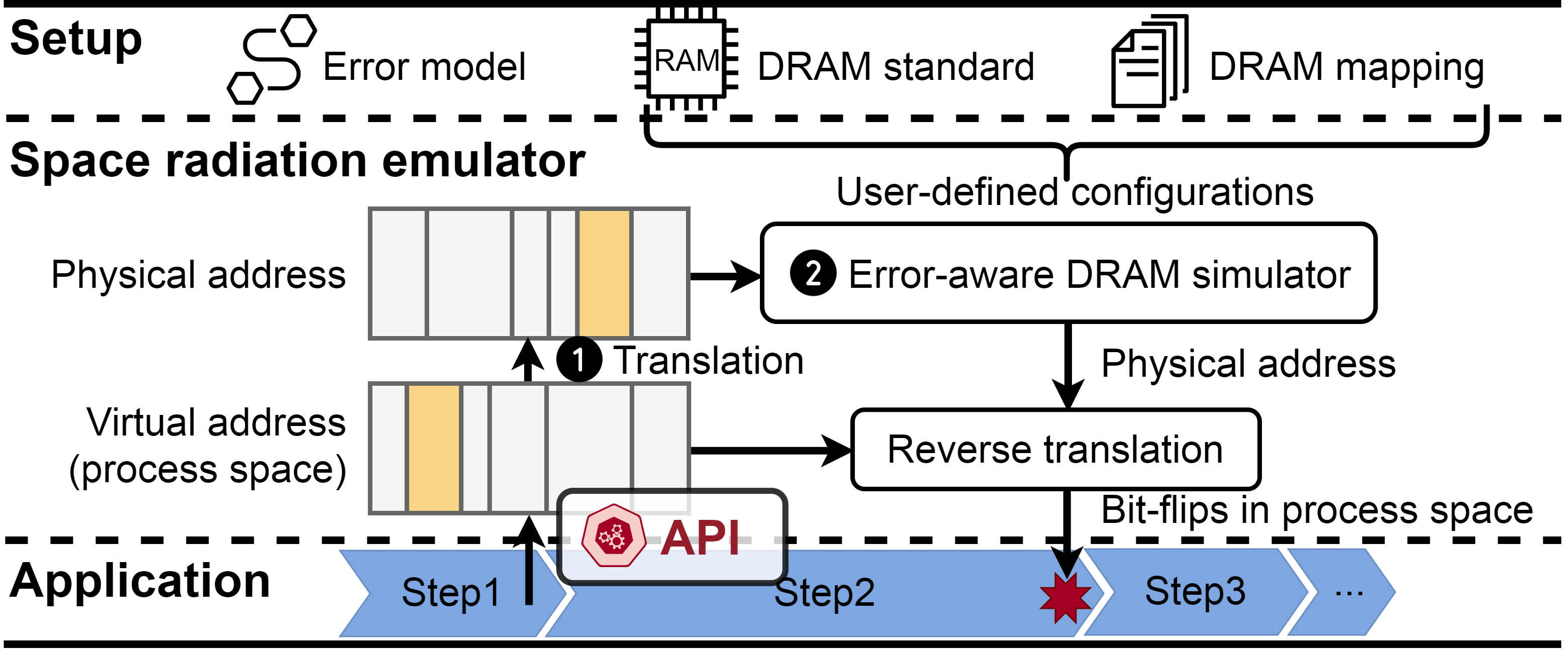}
    \vspace{-1ex}
    \caption{Overview of space radiation error emulator.}
    \label{fig:emulator}
    \vspace{-3ex}
\end{figure}

\paragraphb{{Hardware-in-the-loop space radiation emulator.}}
The prevalence of space radiation-induced bit errors has been repeatedly validated in real satellite environments (see \S\ref{sec:motivation:hazards}--\ref{subsec:radiationThreats}).
To further quantify their impacts on diverse applications at scale,
a flexible and reconfigurable hardware-in-the-loop space radiation emulator is desirable to facilitate both the exploration of applications' error behaviors and the availability of application-aware tolerant design.
Unfortunately, existing DRAM memory simulators~\cite{kim2015ramulator, rosenfeld2011dramsim2, li2020dramsim3} cannot easily meet this demand since they typically delve into the precise cycle in read and write but ignore the on-chip bit error location at the level of memory basic units (i.e., the memory cell that stores one bit).

To this end, we first build an {open-source} hardware-in-the-loop radiation emulator serving for simulating radiation-caused bit errors on diverse executable DNN tasks.
Implementing such an emulator is challenging for three reasons.
First, a real error model of radiation on the memory chip is usually not considered in previous works.
Second, to study the impact of those spatial-correlated memory bit errors from the application perspective, we need an end-to-end memory mapping across multiple layers (see details in Appendix~\ref{appen: mapping}) from radiation-caused hardware bit-flips to corresponding bit errors in applications' runtime process space.
Third, this emulator is supposed to be extensible to various memory configurations and different applications.

We address these challenges as follows. (1) We include the radiation error model as a configurable input of our emulator. 
In this paper, we adopt real-world statistics-based radiation error models validated by~\cite{fabero2020single}. 
(2) We realize the correct and precise error mapping from hardware to application via two steps. 
First, we leverage the Linux kernel interface to translate memory's virtual addresses (handling applications) to physical addresses (\ding{182} in~\figurename~\ref{fig:emulator}). 
Then, we improve the verified classical DRAM simulator~\cite{kim2015ramulator} to map from physical addresses to precise bit locations in the memory chip via the DRAM hierarchy (\ding{183} in~\figurename~\ref{fig:emulator}).
(3) We also expose user-defined memory configuration files and rich APIs for extension.
We pursue our emulator in ease of use: users only need to set up user-defined memory configuration files as required and call the API provided to inject radiation-induced bit errors in their runtime applications. 
In the case of DNN-based applications, we inject bit errors into the runtime DNN models loaded in the memory (i.e., DNN inference engine).
More low-level details of our space radiation emulator are available in Appendix~\ref{appen:remu}.

\begin{figure}[t]
    \centering
    \setlength{\abovecaptionskip}{2pt}
    \setlength{\belowcaptionskip}{2pt}
    \subfloat[Examples of satellite imaging datasets.\label{subfig:tasks_a}]{%
        \begin{minipage}[c]{0.98\linewidth}
            \centering
            \includegraphics[width=\linewidth]{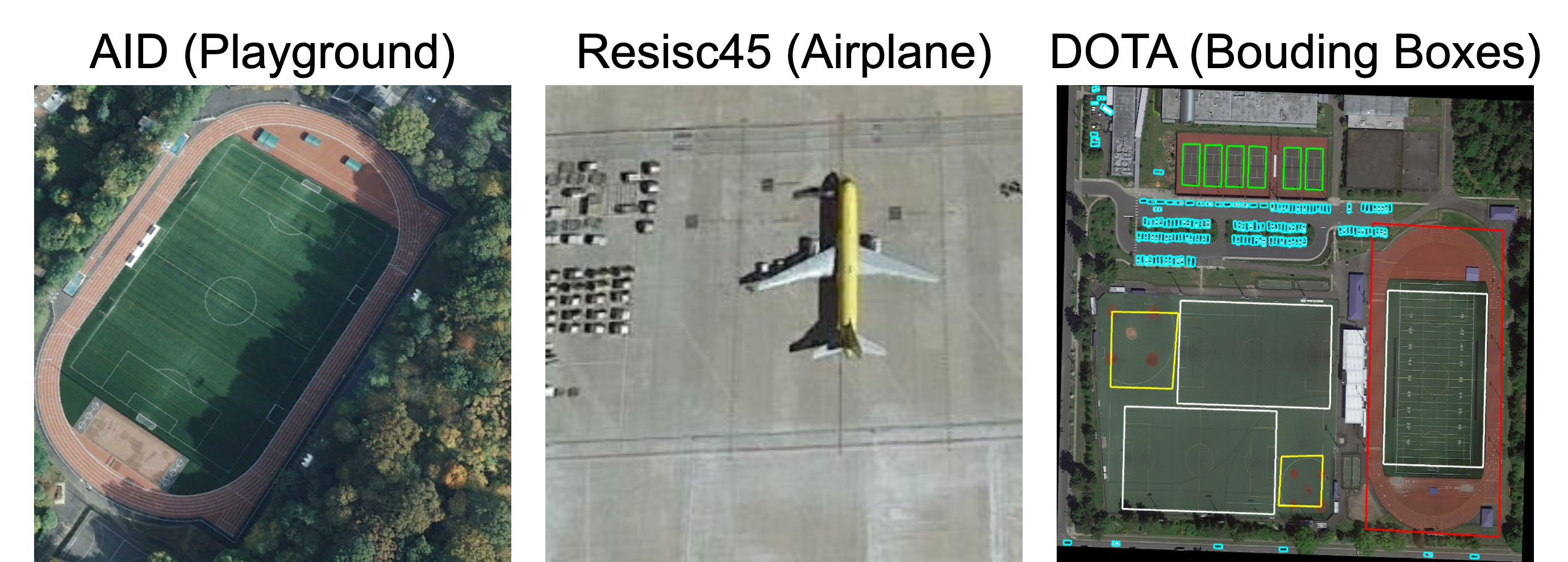}
        \end{minipage}
    }\\ 
    \subfloat[In-orbit DNN tasks considered in this paper.\label{subfig:tasks_b}]{%
        \begin{minipage}[c]{\linewidth}
            \centering
            \resizebox{\linewidth}{!}{
                \begin{tabular}{c|c|c|c|c}
                   \hline
                   \textbf{Task}  & \textbf{Dataset}                 & \textbf{Size, Numbers, Classes} & \textbf{Structure}  & \textbf{Perf.}   \\ 
                   \hline
                   Task1 & \multirow{2}{*}{R45} & {256$\times$256 px, }  & RN50  & 92.0\%     \\
                   Task2 & & 31500 images, 45 classes & DN161  & 95.4\%     \\
                   \hline
                   Task3 & \multirow{2}{*}{AID} & {600$\times$600 px, } & RN50   & 97.4\%     \\
                   Task4 & &  10000 images, 30 classes& DN161  & 85.2\%   \\
                   \hline
                   \multirow{2}{*}{Task5} & \multirow{2}{*}{DOTA} & {800$\times$800-20000$\times$20000 px, }  & \multirow{2}{*}{YOLOv5} & \multirow{2}{*}{76.8\%} \\
                    &  & 2806 images, 15 classes  &  &    \\
                   \hline
                \end{tabular}
            }
        \end{minipage}
    }
    \caption{DNN-based Earth observation datasets and tasks. Perf. means prediction accuracy for Task 1-4 and Average Precision (mAP) for Task5.}
    \vspace{-3ex}
    \label{fig:tasks}
\end{figure}

\paragraphb{{Real satellite imaging datasets.}}
We use three well-known earth observation datasets, i.e., RESISC45~\cite{cheng2017remote} and AID~\cite{xia2017aid} for scene recognition, and employ DOTA~\cite{xia2018dota} for object detection (see examples in \figurename~\ref{subfig:tasks_a}).

\begin{packeditemize}

\item\noindent\textbf{RESISC45 (R45)}~\cite{cheng2017remote} is a large-scale benchmark dataset collected from Google Earth, containing 31,500 images across 45 scene classes with various spatial resolutions.

\item\noindent\textbf{AID}~\cite{xia2017aid} consists of 10,000 multi-resolution images across 30 scene classes collected from Google Earth, featuring from different regions around the world under various imaging conditions and seasons. 

\item\noindent\textbf{DOTA}~\cite{xia2018dota} is a representative object detection dataset collected from multiple sensors and platforms, containing 2,806 images with 188,282 instances across 15 object categories, annotated with oriented bounding boxes.

\end{packeditemize}

\paragraphb{{Representative in-orbit DNN tasks.}}
We show the case of two common DNN-based Earth observation applications in orbit: scene recognition~\cite{esposito2019orbit} and object detection~\cite{wang2022empirical}, which play crucial roles in various tasks such as navigation~\cite{heiselberg2020remote}, environment preservation~\cite{ghazouani2019multi}, real-time disaster detection~\cite{gupta2019xbd, leyva2023satellite}, etc.
Scene recognition aims to classify a given Earth scene into predefined categories, such as harbor, forest, etc. while object detection aims to locate objects of interest in sensory images.
DNN models have inspired the performance in real-time processing of these tasks in orbit~\cite{esposito2019highly, geist2023nasa, denby2023kodan}.

In this paper, we showcase 2 widely-used DNN structures ResNet-50 (RN50)~\cite{he2016deep} and DenseNet-161 (DN161)~\cite{huang2017densely} for scene recognition task, and YOLOv5~\cite{yolov5} for object detection task. 
We list more details of the 5 tasks and their original model performance in Table~\ref{subfig:tasks_b}. 
All 5 tasks' DNN models are trained on a ground server (details in Appendix~\ref{appen:training}) and deployed on the real satellite payload (\S~\ref{subsubsec:scenario}).

\subsection{Applications' Sensitivity to Radiation}
\label{subsec:senanaly} 

\begin{figure}[!htbp]
    \centering
    \includegraphics[width=1\linewidth]{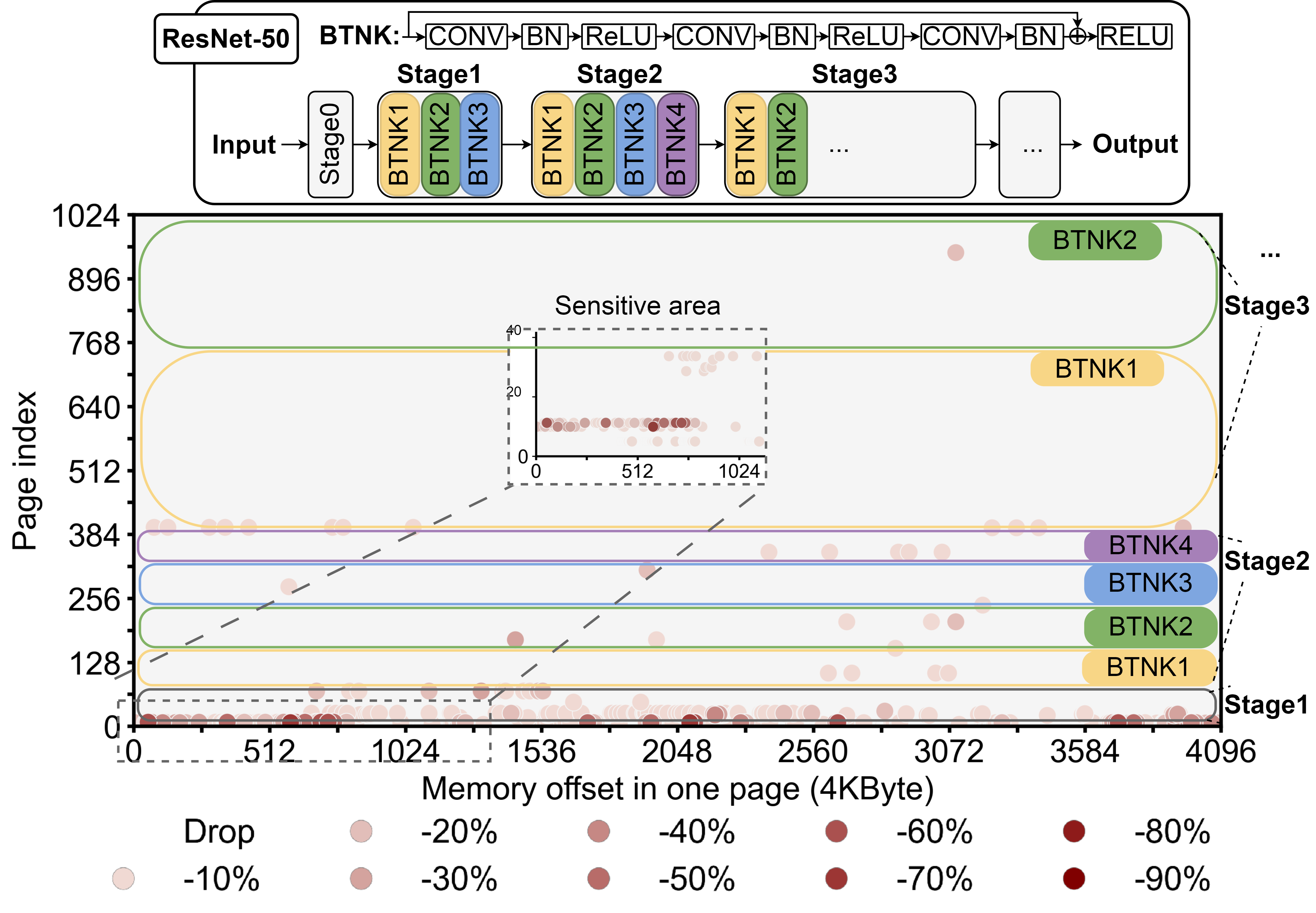}
    \vspace{-4ex}
    \caption{
    Radiation-induced bit error sensitivity in Task 1.
    }
    \label{fig:task1loop}
    \vspace{-4ex}
\end{figure}

To disclose opportunities for application-aware space radiation protection,
we analyze the runtime DNN inference's sensitivity to radiation-induced bit errors.  
Specifically, we scan the whole memory space storing the DNN inference engine (see details in \ref{subsubsec:scenario}) by using our emulator for one-by-one bit-flipping to see how single-bit errors influence the models' accuracy on their testing dataset. 
Different from previous works that target static DNN parameter errors, we focus on the DNN runtime stage by analyzing the engine file loaded in memory. 
Please also note that our DNN engine file, memory configuration (used by the satellite payload system), and radiation-caused error models are all configurable inputs such that our work can be extended to other tasks as well.
By including the memory configuration as well, we comprehensively analyze the sensitivity of \textit{every bit} in one runtime DNN model in memory (\figurename~\ref{fig:DNNdataflow}) against the bit error on a real satellite's payload system.

As illustrated in \figurename~\ref{fig:task1loop}, we take Task 1 (RN50 model) as an example to display the bit error sensitivity of the runtime DNN model in memory. 
We present the DNN model's memory space with a two-dimensional format by listing each memory page (containing 4K Byte) along the X-axis and putting 1K pages along the Y-axis. 
This format aligns with the real layout in the payload's memory. 
We mark those sensitive bits inside the memory space. 
Note that the RN50 model consists of many layers separated by bottlenecks (BTNK) arranged in several stages in memory. 
For instance, the second stage has four bottlenecks, each consisting of ordered CONV, BN, and ReLU layers as shown in~\figurename~\ref{fig:task1loop}.
We map these layers to the allocated memory space at bottleneck granularity. 

\begin{figure}[!htbp]
    \centering
    \includegraphics[width=\linewidth]{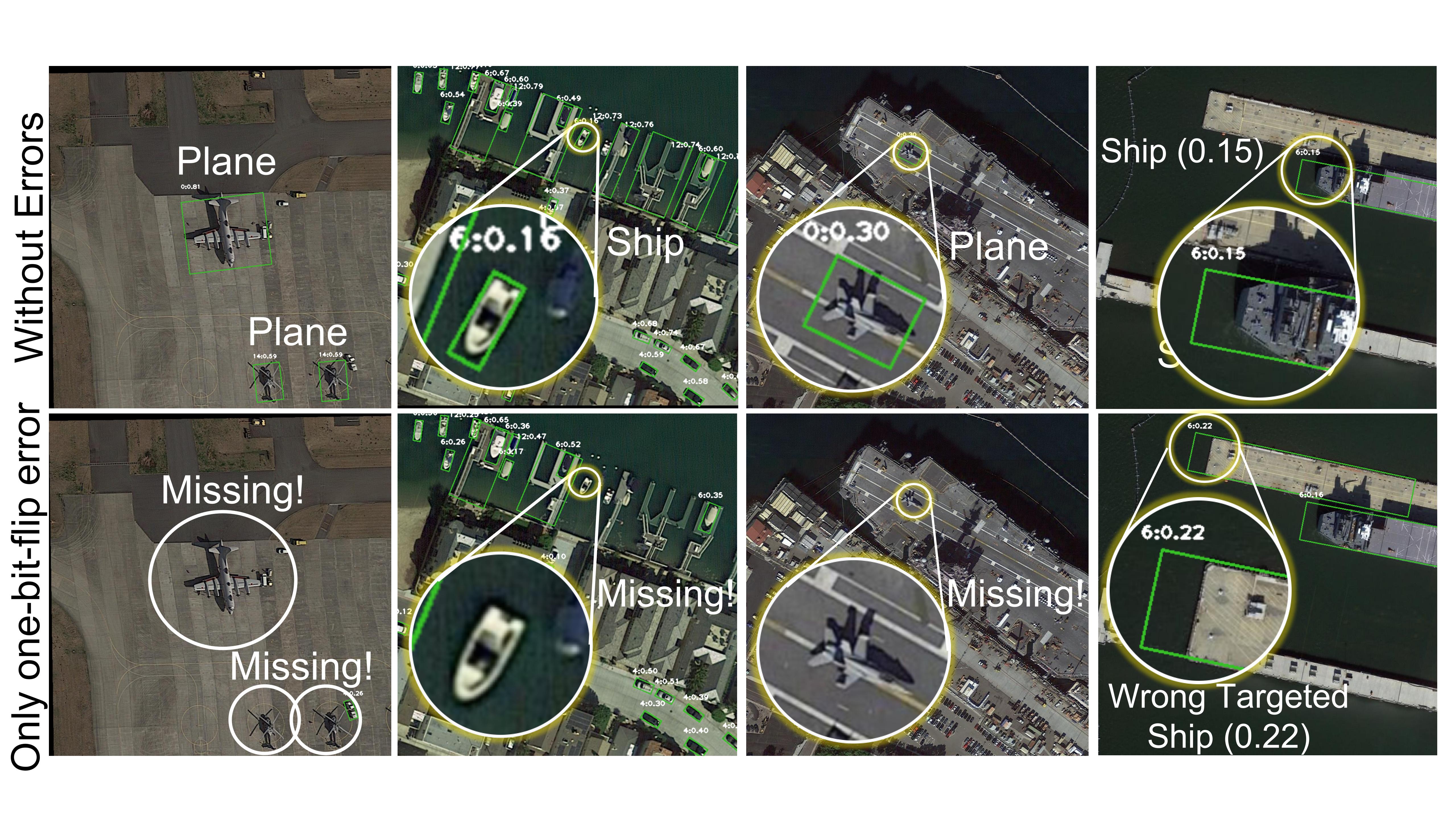}
    \vspace{-4ex}
    \caption{1-bit error in Task 5 (YOLO) leads to wrong results.}
    \label{fig:yoloerror}
    \vspace{-3ex}
\end{figure}

Besides, we also show the influence of such bit errors on the DNN model's results using Task5 in \figurename~\ref{fig:yoloerror} which only 1 critical bit error could lead to objects missing or mis-detected.

\subsection{Uneven Space Radiation Sensitivity}
\label{subsec:unevensensitive}

We find that not all radiation-induced bit errors are equal for applications.
Due to in-orbit tasks' varying sensitivity to different data value inputs and control logic, some bit errors are much more critical than others for applications.
This offers opportunities to adapt the radiation protection cost to the application-specific bit error sensitivity diversity.

Specifically, in the context of in-orbit DNN tasks, 
we find that {\bf {\em shallow} DNN layers are more sensitive to bit errors.}
Radiation-sensitive bits in the DNN inference are unevenly distributed layer-wise, mainly concentrated in two parts, i.e., the very first few layers and some layers in the middle. 
As shown in \figurename~\ref{fig:task1loop}, DNN tasks' sensitive bits distribution concentrates in certain DNN layers (i.e., mainly initial shallow DNN layers and a few middle DNN layers). 
Different DNN models have different sensitivity areas in memory, e.g., the DNN model for Task 2 has more sensitivity bits than Task 1.

\paragraphb{Root cause: Bit error propagation across DNN layers.}
First, DNN inference is computed layer by layer such that the input value of the $(i+1)$-th DNN layer comes from the output value of the $i$-th DNN layer. 
Therefore, once bit errors in the initial shallow DNN layer's parameters occur, their influence on this layer's intermediate calculation results can easily be enlarged in the following DNN layer's calculation during the DNN inference's forward propagation. 
This copes with other works that point out the accurate parameters in shallow layers are more valuable in the model decision~\cite{rakin2019bit, wang2023aegis}. 

Second, such error propagation can be aggravated by the inherent redundancy (i.e., more and more DNN layers) of the modern DNN model itself.
Generally, it is meaningful in the deep learning domain to build deeper neural network models that can fit more complex data and achieve better prediction accuracy in total. 
It is necessary for practitioners to use DNN models with more DNN layers to finish their tasks better. 
However, considering the influence of radiation-induced bit errors, the redundant computation complexity with deep layers will reduce reliability due to potential error propagation and enlargement.

\paragraphb{Design implication 1:}
{\em By distinguishing (in)sensitive bit errors and handling them differently
with domain knowledge, it is possible to protect applications against space radiations
with fewer hardware/software redundancies at lower costs.}

\subsection{{Spatially Correlated Multi-bit Errors}}
\label{subsec:seuvsmcu}

\begin{table}[!htbp]
\centering
\resizebox{0.95\linewidth}{!}{
\begin{tabular}{c|c}
\hline
\textbf{Bit error type} & \textbf{Accuracy degradation} \\ \hline
100 bit error (MCUs+SEUs) & -47.8\%$\sim$-96.1\% \\ \hline
100 bit error (SEUs) & -31.3\%$\sim$-56.2\% \\ \hline
100 bit error (dumb random~\cite{li2020defending}) & -0.9\%$\sim$-1.4\% \\ \hline
\end{tabular}
}\vspace{-1ex}
\caption{DNN inference accuracy degradation when injecting 100-bit space radiation-induced errors for 1,000 times. 
Runtime DNN models (Task 1) in memory are more vulnerable to spatially correlated bit errors (caused by MCUs) than the same number of single-bit errors (consider only SEUs).}
\label{table: remu and random}
\vspace{-3ex}
\end{table}

As introduced in \S\ref{subsec:radiationThreats}, in-orbit computing suffers from both single-bit errors and (spatially correlated) multi-bit errors by space radiations. 
In the context of DNN-based tasks, we find that {\bf spatially correlated multi-bit errors are more detrimental to runtime DNN models.} 
\tablename~\ref{table: remu and random} quantifies their degradations of the DNN model's accuracy by setting the same number of bit errors (i.e., 100 bits) but the real radiation error models (details in Appendix \ref{subsec: configurations}) are pure SEUs and SEUs+MCUs, respectively. 
It shows that multi-bit errors (MCUs) degrade more DNN inference accuracy than radiation-induced single-bit errors (SEUs) and randomly bit-flip errors in DNN models without hardware-in-the-loop~\cite{chen2021low, hoang2020ft, ning2021ftt, li2020defending} (denoted as ``dumb random'' in \tablename~\ref{table: remu and random}).

\paragraphb{Root cause: Spatial bit error correlation.}
The reason for the greater impact of correlated multi-bit errors on DNN is twofold.
First, observing runtime DNN models in memory space (\figurename~\ref{fig:loop} and \figurename~\ref{fig:task1loop}), those sensitive bits are clustered together. 
Once those bits and their spatial-correlated bits are flipped together (i.e., MCU effects), there will be a significant error inside their DNN layer's calculation. 
Second, even though most bits in the memory space are non-sensitive, a cluster of adjacent bits that flip together in arbitrary memory space can also lead to more serious accuracy degradation than the same number of single-bit errors that are randomly distributed inside the memory space. 
This copes with the experiment results in \tablename~\ref{table: remu and random}. 
Note that spatially correlated multi-bit errors cannot be simplified as the same number of arbitrary random one-bit errors. 
Therefore, although previous works have studied the DNN model's sensitivity against random bit errors, their solutions cannot be adopted for tackling spatially correlated multi-bit errors in space.

\noindent\textbf{Design implication 2:} 
{\em For cost efficiency, it is helpful to tackle the spatial correlation among radiation-induced multi-bit errors rather than treat them separately as independent single-bit errors.}

\section{Design of \name}
\label{sec:design}

Our findings in \S\ref{sec:dnnanalysis} imply that application-aware space radiation protections may be a more cost-effective paradigm for in-orbit computing.
We make such a case in the context of satellite DNN tasks by designing
\name~(\underline{R}adiation \underline{E}rror-tolerant \underline{D}eep neural \underline{Net}work) to achieve three goals:

\begin{packeditemize}
\item \textbf{Radiation-tolerance:} \name~should restrain radiation-induced errors from degrading the DNN model accuracy;

\item \textbf{Cost-efficiency:} 
\name~should retain low satellite resource overhead compared to existing protections in \S\ref{subsec:classicalProtections}.

\item \textbf{Extensibility:}  \name~should be applicable to different DNN-based satellite tasks while achieving the above goals.

\end{packeditemize}

\subsection{Design Overview} 

The key ideas of \name~are twofold.
First, instead of trying to correct arbitrary bit errors in a DNN model, we aim to \textit{suppress the error propagation and enlargement layer by layer} from initial shallow DNN layers to the deeper DNN layers. 
Second, we add the \textit{model-level redundancy for decision} by allowing the model to decide dynamically, i.e., allowing input samples to be predicted at arbitrary DNN layers when a certain confidence threshold is met. 
This can mitigate the influence of a set of spatially correlated bit errors in one certain area (e.g., MCUs in sensitivity areas in the middle layers). 
Note that the above two intuitions can roughly be applied to \textit{any} DNN models which are built with many layers connected by activation functions.

\begin{figure*}[t]
    \centering
    \includegraphics[width=0.98\textwidth]{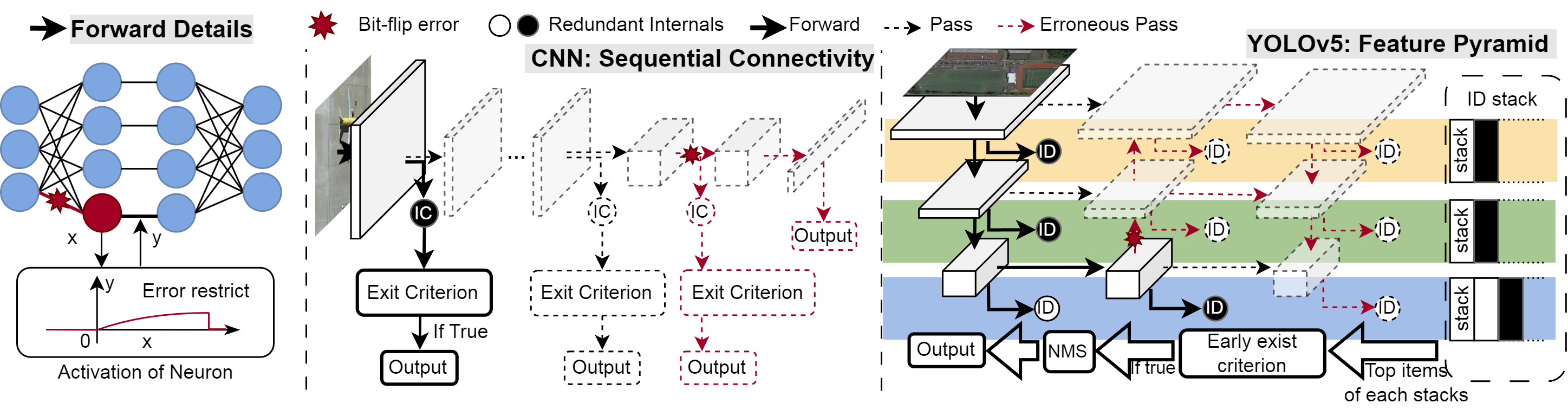}
    \vspace{-1ex}
    \caption{Overview of \name for space radiation-tolerant in-orbit DNN tasks.}
    \label{fig:redOverview}
    \vspace{-3ex}
\end{figure*}

\figurename~\ref{fig:redOverview} overviews how \name~renovates a satellite DNN model to tolerate space radiations.
It takes two steps:

{\bf (1) Bounding errors in each DNN node (\S\ref{subsec: log}):} 
For each neuron node in the DNN model, we replace the most widely-used activation function (ReLU) with a logarithmic activation function to suppress error propagation and amplification.  
This is motivated by the fact that the classical ReLU activation function's output is infinitely large when its input is infinitely large, thus being easy to amplify erroneous inputs {\em unboundedly}. 
Instead, \name~uses the activation function that has an upper bound on the output value to prevent the amplification of wrong inputs, and its output growth gradually slows down as the input increases. 
By replacing this specific activation function in each layer of one DNN model, \name~can suppress the bit error caused by wrong output's propagation and enlargement layer by layer. 

{\bf (2) Suppressing error propagation across DNN layers (\S\ref{sec:sdn}):}
To mitigate radiation-induced error propagation across DNN layers, \name~adopts a decision-redundant strategy for the DNN model's inference. 
We allow the DNN model to stop inference early {\em before} traversing all its neural layers if it is confident enough with its current outcome.
The rationale is to let some easier input samples be predicted at shallow layers only, thus avoiding error propagation to deeper layers.

\subsection{Robust Logarithmic Neural Activation}
\label{subsec: log}

\begin{figure}[!htbp]
    \centering
    \includegraphics[width=0.98\linewidth]{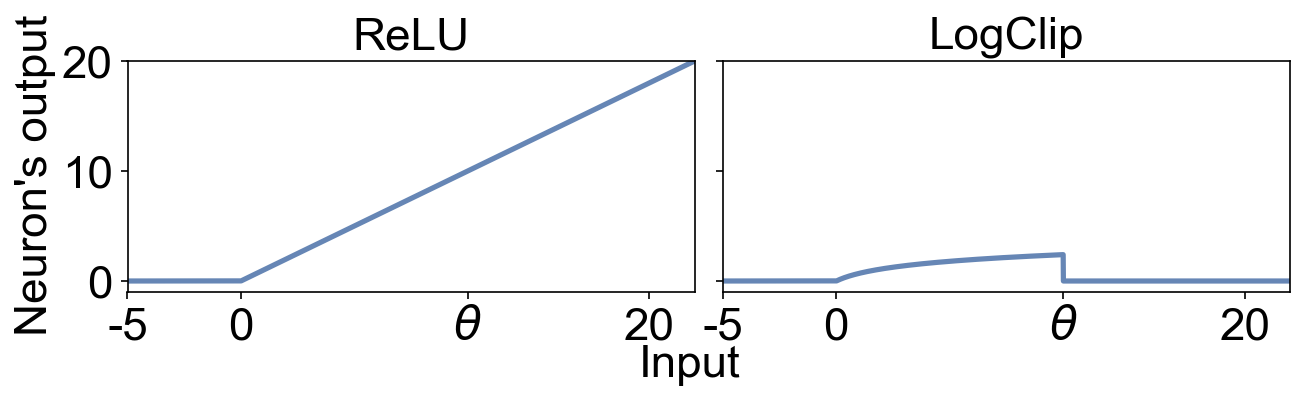}
    \vspace{-2ex}
    \caption{ReLU and \name's activation function.}
    \label{fig:activation-showcase}
    \vspace{-3ex}
\end{figure}

\name~uses a non-linear and slow-growing activation function to replace the classical ReLU activation function~\cite{agarap2018deep} in DNN models.
{As shown in \figurename~\ref{fig:activation-showcase},} the existing DNN models' ReLU activation function's output is infinitely large when its input is infinitely large, thus being easy to amplify erroneous inputs {\em unboundedly}. 
Instead, we note that logarithmic functions naturally constrain positive infinity and address linear growth with more gradual logarithmic growth. 
We clip the negative growth of logarithmic functions such that if the input is less than 0, the output is 0
We note that using $\mathrm{Log}$ activation to train the model converges quickly which does not introduce extra cost during model training compared with using ReLU.
Like~\cite{chen2021low, hoang2020ft}, we further limit the positive growth by clipping the output greater than the maximum bound $\theta$ to 0.
The maximum bound $\theta$ is calibrated in the post-training phase, using calibrating datasets (i.e., partial of the training set) to determine the maximum output value for each layer. 
The activation function $\mathrm{LogClip}$ is defined in Eq.~\ref{eq:logclipact}.

\vspace{-2ex}
\begin{equation}
    \label{eq:logclipact}
    \mathrm{LogClip}(x) = 
    \begin{cases}
        0 & \text{if } x \leq 0 \text{ or } x > \theta, \\
        \log(x + 1) & \text{if } 0 < x \leq \theta.
    \end{cases}
\end{equation}

\begin{figure}[!htbp]
    \centering
    \includegraphics[width=0.96\linewidth]{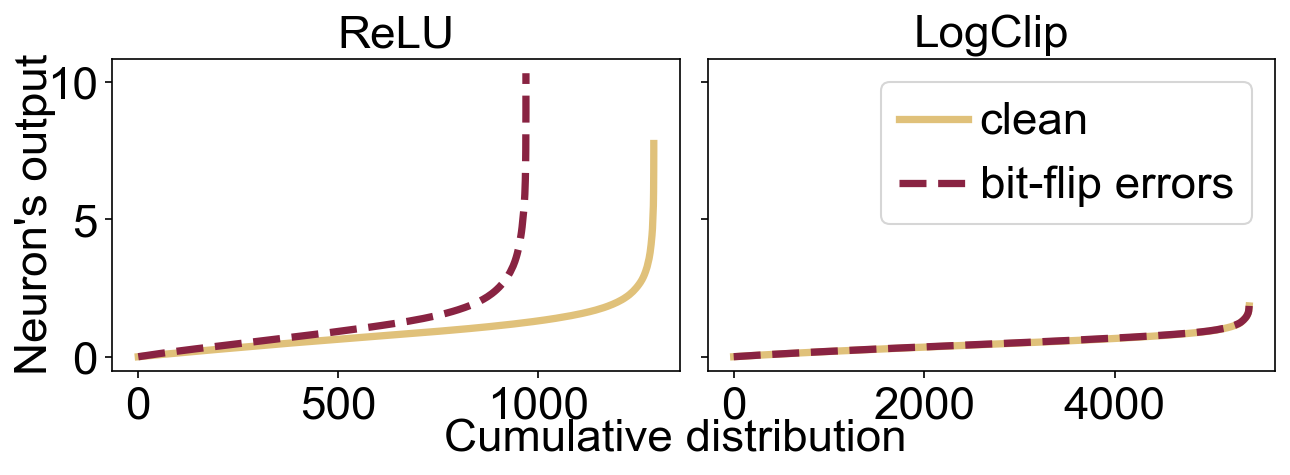}
    \vspace{-2ex}
    \caption{Distributions of intermediate DNN layer's outputs against bit errors of ReLU and our activation function.}
    \label{fig:cdf}
    \vspace{-3ex}
\end{figure}

\figurename~\ref{fig:cdf} compares the error-suppressing ability of our $\mathrm{LogClip}$ activation.
We use the cumulative distribution~\cite{drew2000computing} to characterize the output values' distribution of one DNN layer.
We inject errors in some bits before this layer and observe how the output is influenced by those injected bit errors. 
For instance, in \figurename~\ref{fig:cdf}, we observe the deviation of the red line from the yellow line.
We find a greater deviation in using the ReLU activation function than $\mathrm{LogClip}$.
Thus, using an activation function with \textit{a clear upper bound on output and which becomes gradually less sensitive to increases in input} can effectively suppress the error propagation. 

\subsection{Multi-Exit Structure across DNN Layers}
\label{sec:sdn}

Across the DNN layers, \name~introduces a multi-exit (ME) structure to mitigate the error effect. 
The ME structure was initially proposed to solve the model's overthinking issue~\cite{huang2017densely}, which allows simpler samples to be predicted by shallow layers with various internal classifiers (ICs).
Instead, \name's ME structure is designed to avoid radiation-induced error propagation and amplification.
We improve this structure to let partial samples exit early to avoid potential errors in deeper layers. 
Also, ME can effectively reduce the average number of DNN layers that all samples pass through to make inference more lightweight. 
\name~offers two types of ME designs for different DNN-based satellite imaging tasks:

\paragraphb{Scene recognition tasks.}
For the scene recognition tasks based on convolutional neural network (CNN) models like our Task 1 to Task 4, we adopt existing ME structures.  
For each model, we define that the model $G$ has N+1 layers containing N hidden layers and one final output layer. 
To build ME model $\hat{G}$, we attach an internal classifier (IC) for each hidden layer, denoted as $I_l$, $(1<=l<=N)$, which typically consists of a convolutional layer, a batch normalization layer, and a fully connected layer as shown in~\figurename~\ref{fig:redOverview}. 
These ICs can let simpler samples with enough confidence scores (i.e., greater than the predefined exit threshold $T$) get predicted results and stop the inference. 
Thus, as shown in left in \figurename~\ref{fig:redOverview}, ME on our CNN-based model can allow partial samples to be decided at arbitrary DNN layer which can avoid significant errors (usually caused by MCUs) in deeper layers.

\noindent\textbf{Object detection tasks.}
Unlike scene recognition, exiting ME structure is hard to directly adopt on another vital task in-orbit, object detection (e.g., our Task 5 based on YOLO~\cite{redmon2018yolov3}). 
The main reason is models like YOLO have complicated feature pyramid networks~\cite{lin2017feature}~and an uncertain number of detection boxes (in contrast to a certain number of classes of models for scene recognition tasks). 

To mitigate radiation-induced bit errors, we design a new decision-redundant DNN structure, especially for object detection tasks based on models like YOLO. 
We notice that unlike the commonly sequential connected layers of DNN models like RN50, YOLO employs a network structure containing hierarchical and multi-scale feature aggregation (e.g., three feature scales in YOLOv5~\cite{yolov5}). 
Thus, instead of directly adopting internal classifiers on sequential DNN layers, we design \textit{internal detectors} (IDs in \figurename~\ref{fig:redOverview}) targeting feature maps at three scales on YOLOv5. 
We attach $N$ IDs at various intermediate layers of YOLOv5 with different feature scales, denoted as $I_l$ $(1<=l<=N)$ as illustrated in ~\figurename~\ref{fig:redOverview}. 
To efficiently handle the outputs of these multi-scale IDs, we maintain a stack where new detection results are pushed as they are generated for each feature scale. 
The decision to potentially stop the inference early is made by evaluating an exit criterion based on the latest detection results from each scale. 
This early exit criterion is defined as follows:
{\setlength{\abovedisplayskip}{5pt} 
\setlength{\belowdisplayskip}{5pt}
\begin{flalign}
\frac{1}{|S|} \sum_{i \in S} (o_{i} \cdot c_{i}) > T,\quad S = \{i \mid o_i > \theta,\ i = 1\ , \ldots,\ M\} 
\end{flalign}}
where $M$ denotes the total number of predicted boxes (without non-maximal suppression (NMS)), $o_i$ is the probability of the presence of an object, $\theta$ represents the criterion for object presence, $c_i$ reflects conditioned on the grid cell containing an object~\cite{redmon2018yolov3}~, and $T$ is the predefined threshold for early exit.

\paragraphb{Error-tolerant model training:} 
We last discuss how to train the above DNN models to enable the full-fledged \name~design.
We first train a model $G$ with our activation functions following a typical DNN training process. 
Then, we insert $N$ ICs/IDs for this model $G$ and fine-tune to get the protected model $\hat{G}$. 
The loss function of training $\hat{G}$ is as follows. 
\vspace{-1ex}
\begin{equation}
\label{eq:branch_loss}
\mathcal{L}_{IC/ID}=\sum_{l=1}^{N}{\sum_{\left( x_i,y_i \right) \in \mathcal{X}}{\mathcal{L}\left( f_l\left( x_i \right) ,y_i \right)}}, 
\vspace{-1ex}
\end{equation}
where $f_l\left( x_i \right)$ denotes the output of $I_l$, $\mathcal{X}$ denotes the training datasets we use to build the ME model, and $\mathcal{L}$ is a typical training loss function of $G$.
More training details are in~\ref{appen:training}.

\paragraphb{Solution analysis:}
\name~achieves all its goals as follows:

\begin{packeditemize}
\item {\bf Radiation-tolerance:} \name~can tolerant radiation-caused bit errors by suppressing shallow DNN layer's error propagation and avoiding a cluster of spatially correlated bit errors in the middle or deeper DNN layers. 

\item {\bf Cost-efficiency:} \name~only requires a tiny extra memory footprint (for ICs or IDs) at runtime but can significantly accelerate the inference speed and reduce energy consumption which is crucial for orbital computing (see results in \S\ref{sec:eval}).
The core designs of \name~targets on enhancing the fundamental DNN structures. 

\item {\bf Extensibility:}
The core designs of \name~targets on enhancing the fundamental DNN structures.  
For different orbital computing tasks (i.e., scene recognition and object detection), and can be further extended on other DNN-based orbital computing tasks as well. 
\end{packeditemize}

\section{{Evaluation}}
\label{sec:eval}

This section first experiments \name~in the real-world end-to-end ground testing used by Chaohu-1 satellites to validate \name's functionality, accuracy, and cost efficiency in real orbital computing settings (\S\ref{sec:eval:real-test}).
Then, we perform hardware-in-the-loop emulations at scale to delve into how \name~can tolerate various radiation-caused bit errors (\S\ref{sec:eval:emulation}).

\subsection{{Real-World End-to-End Ground Testing}}
\label{sec:eval:real-test}

We experiment \name~in the real-world end-to-end ground testing on the Jetson Xavier NX platform~\cite{jetson_xavier_nx} which is deployed as a payload on Chaohu-1 SAR satellite~\cite{Chaohu-1} launched in 2022. 
We list test settings in \S~\ref{subsubsec:scenario} and results in \S~\ref{subsubsec:results}.

\subsubsection{Ground Test Setup}
\label{subsubsec:scenario}

\vspace{-2ex}
\begin{figure}[!htbp]
\subfloat[Orbital computing includes 3 steps. \label{subfig:groundtest_a}]{
    \begin{minipage}{0.6\linewidth}
        \centering
        \includegraphics[width=\linewidth]{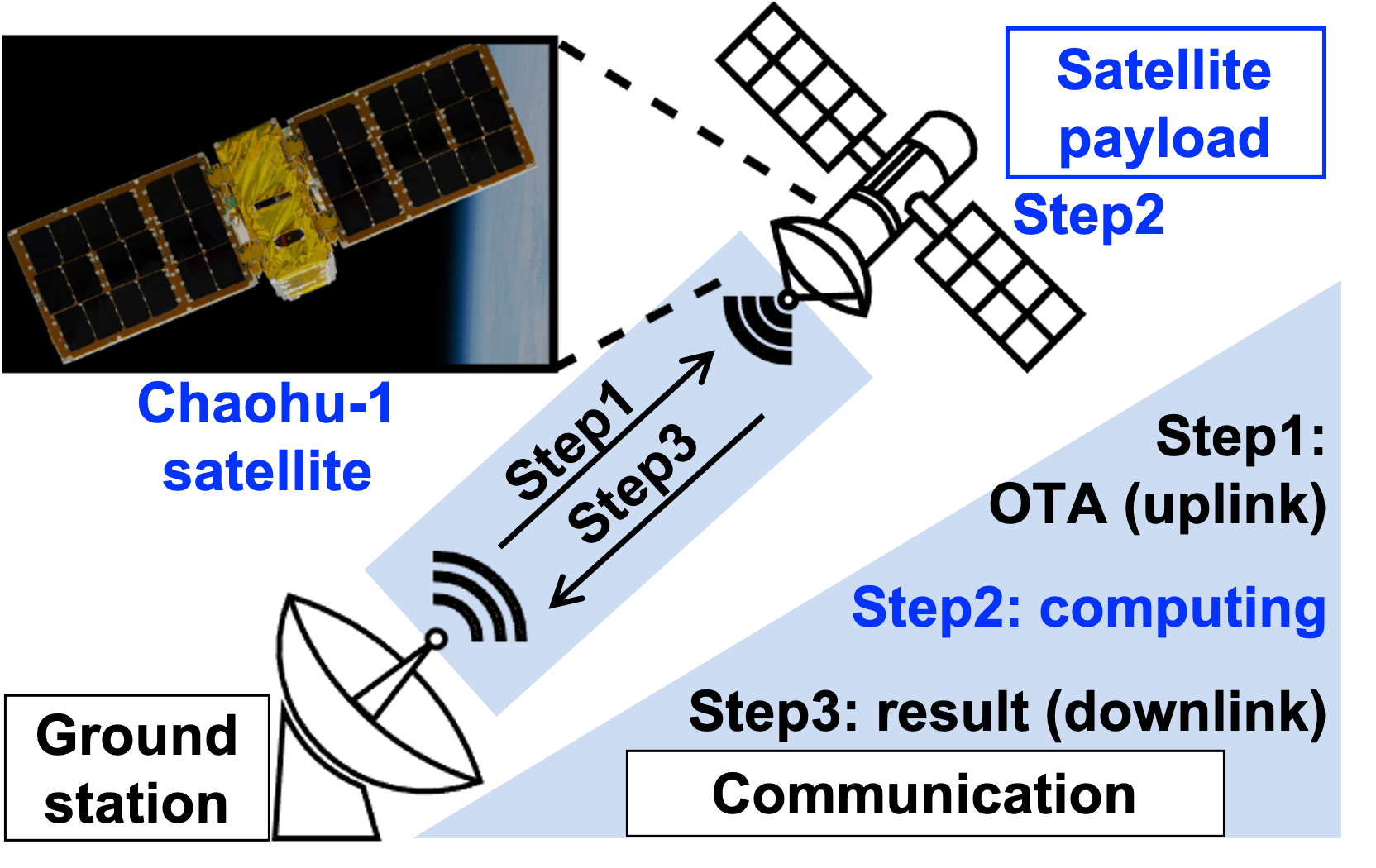}
    \end{minipage}
    }
\subfloat[Test environments.\label{subfig:groundtest_b}]{
    \begin{minipage}{0.35\linewidth}
        \centering
        \includegraphics[width=\linewidth]{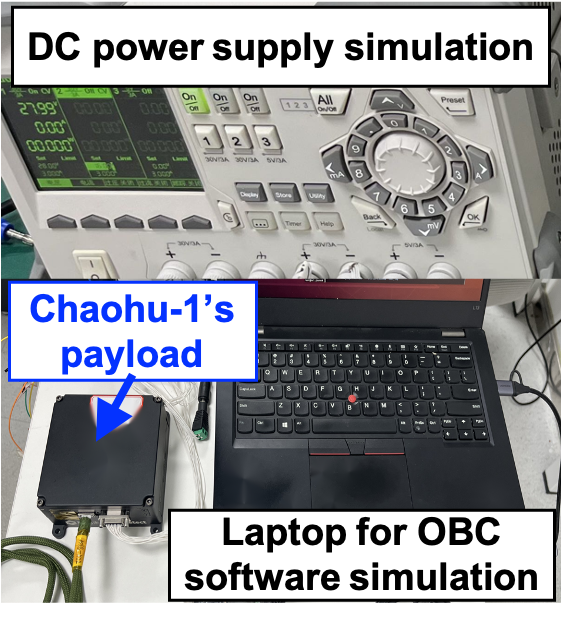}
    \end{minipage}
    }
    \vspace{-2ex}
    \caption{We use \textbf{Chaohu-1's satellite payload} with its OBC software for our end-to-end ground test.}
    \label{fig:crash}
    \vspace{-2ex}
\end{figure}

We use Chaohu-1's satellite payload to test \name's functionality and effectiveness.
As illustrated in \figurename~\ref{subfig:groundtest_a}, a real in-orbit computing test typically takes three steps:

{\em (1) Over-the-air (OTA) application loading into satellites:} 
Chaohu-1 satellite supports a high-speed uplink of up to 1Mbps for the OTA upgrade. 
Taking deploying DNN models in orbit as an example, both the packed payload-compatible DNN engine file (e.g., $\sim$20MB for Task 5 with \name) and its related programs for execution in-orbit ($<$1MB) need to be upgraded on the satellite's on-board computer (OBC). 
This OTA upgrade can only be performed when communication is available between the ground station and the Chaohu-1 satellite. 
In fact, this will require 2-3 tracks (each track takes around 90 minutes) to finish with transmission breakpoint resumption and onboard splicing. 

{\em (2) In-orbit computing execution:}
After OTA upgrades, the OBC will hold until receiving the corresponding command code.
Then, the OBC will power on the payload and synchronize the time (to guarantee accurate timestamp with results) and then deploy the application on the payload for execution. 

{\em (3) Result collection:}
After the payload finishes the computation (e.g., inference for given samples), the OBC will collect the results from the payload and send the results back to the ground station for post-analysis.

In practice, the above real in-orbit computing test procedure is time-consuming and inflexible.
Instead, before conducting these real tests, satellite operators (e.g., Chaohu-1) usually perform intensive testing with real-world satellite payload systems on the ground. 
Following Chaohu-1's real payload and ground test procedures, 
we build the test environment in \figurename~\ref{subfig:groundtest_b}. 
The core device, satellite computing payload, is identical to the one deployed on the Chaohu-1 satellite. 
This payload consists of Jetson Xavier NX~\cite{jetson_xavier_nx} and an FPGA chip, where NX is used for onboard DNN inference, and FPGA is used as the interface between the NX and the OBC.
Then, the OBC is simulated by software on the laptop, and the power of the payload is simulated by a 16V DC power supply.
We follow the above typical orbital computing procedure to test.

To deploy the application (e.g., DNN models) into orbit, a prerequisite step is to convert and deploy it compatible with payload systems. 
We use TensorRT 8.2.1 SDK~\cite{tensorrt}, which is compatible with our satellite payload systems (i.e., Jetpack 4.6.1, CUDA 10.2, cuDNN 8.2.1), to convert and deploy the 5 tasks' trained models for onboard inference. 
Model conversion and deployment are given as follows.
First, for model conversion (still offline), TensorRT creates and serializes an optimized DNN engine file converted from the trained DNN model. 
Note that when creating the DNN engine file we perform the post-training quantizaion~\cite{tensorrt} from floating-point precision to INT8 precision (i.e., quantize both parameters and activations from 32-bit floating-points to 8-bit integers). 
This step is typically required in the onboard deployment, which can achieve faster inference speed and lower memory footprint for inference in orbit. 
Second, for model deployment (online), the DNN engine file is deserialized into the memory\footnote{On the satellite's payload system (i.e., Jetson Xavier NX), the memory used by GPU and CPU is physically unified (i.e., only one memory chip), so DNN is allocated in memory once for execution. No {data} transfer between main memory and GPU memory is needed.} along with all input and output allocation in preparation for runtime inference executions. 
More settings of \name's runtime inference acceleration are available in Appendix~\ref{appen:speed}.

\subsubsection{Results}
\label{subsubsec:results}

Table~\ref{tab:combined_efficiency} compares \name~with the legacy DNN models without early exits in \S\ref{sec:sdn} in Task 1 and 5. 
In this table, the inference speed per sample indicates the task's actual calculation speed on the payload. 
The other metric, {throughput, is more important in orbital computing since this metric determines whether this in-orbit DNN can process the data generated by the satellite's camera}. 
Most of today's SAR satellites equip high-definition cameras that can take photos and generate earth observation images at a high speed (e.g.,
Chaohu-1's camera can generate earth observation images at a speed of at least 100Mbps~\cite{pri2024}). 
This means the deployed DNN in-orbit must be able to process at least this speed to meet the requirement of real-time orbital computing.  

\begin{table}[t]
\centering

\begin{subtable}{0.98\linewidth}
\centering
\resizebox{\linewidth}{!}{
\begin{tabular}{c|cccccc}
    \hline
    \textbf{Exit Threshold}      & \textbf{0.6} & \textbf{0.7} & \textbf{0.8} & \textbf{0.9} & \textbf{0.99} & \textbf{1.00} \\
    \hline
    \textbf{Accuracy}                 & 79.4 & 85.4 & 89.6 & 92.4 & 94 & 94 \\
    \hline
    {\textbf{Throughput}}    & 124.1 & 121.1 & 116.7 & 112.0 & 103.9 & 95.8 \\
    \textbf{(Mbps)}                                     & +29.6\% & +26.4\%  & +21.8\%  & +16.9\% & +8.4\% & +0.0\% \\
    \hline
    \textbf{Inference speed} & 5.2 & 6.2 & 7.2 &  8.2 & 10.2 & 12.3 \\
    \textbf{per sample (ms)}                                    & -57.7\% & -49.3\% & -41.5\% & -33.5\% & -17.3\% & 0.0\% \\
    \hline
\end{tabular}
}\vspace{-1ex}
\caption{Efficiency improvement in Task 1.}
\label{tab:RN_R45_different_exit_threshold} 
\vspace{1ex}
\end{subtable}

\begin{subtable}{0.98\linewidth}
\centering
\resizebox{\linewidth}{!}{
\begin{tabular}{c|cccccc}
    \hline
    \textbf{Exit Threshold}      & \textbf{0.10} & \textbf{0.20} & \textbf{0.30} & \textbf{0.40} & \textbf{0.45} & \textbf{1.00} \\
    \hline
    \textbf{mAP}                 & 76.5 & 78.3 & 80.3 & 78.6 & 78.8 & 79.7 \\
    \hline
    \textbf{Throughput}    & 188.3 & 172.7 & 158.1 & 146.2 & 140.8 & 133.9\\
    \textbf{(Mbps)}                                     & +40.6\% & +29.0\% & +18.1\% & +9.2\% & +5.2\% & +0.0\%\\
    \hline
    \textbf{Inference speed} & 112.5 & 126.4 & 142.1 & 161.5 & 172.2 & 188.0\\
    \textbf{per sample (ms)}                                    & -40.2\% & -32.8\% & -24.4\% & -14.1\% & -8.4\% & 0.0\%\\
    \hline
\end{tabular}
}\vspace{-1ex}
\caption{Efficiency improvement in Task 5.}
\label{tab:yolo_different_exit_threshold}
\end{subtable}
\vspace{-2ex}
\captionof{table}{
\name's tunable inference speed for Task 1 and Task 5 on Chaohu-1's payload. 
}\vspace{-2ex}
\label{tab:combined_efficiency}
\end{table}

For the multi-exit structure, we can set different thresholds of early exit for \name. 
This can effectively improve the throughput and reduce the inference time of GPU. 
For instance, for Task 1, when the threshold is set as 0.99, the accuracy is not dropped, but the throughput is improved by 8.4\% due to some samples' early exit, which can catch up with the speed of Chaohu-1's camera to achieve a real-time decision.
We indicate the ability of \name~to balance the trade-off between efficiency (i.e., throughput) and performance (i.e., accuracy).
Thus, \name~can be used as a cost-aware application in orbit. 

\begin{figure}[!htbp]
    \centering
    \includegraphics[width=0.9\linewidth]{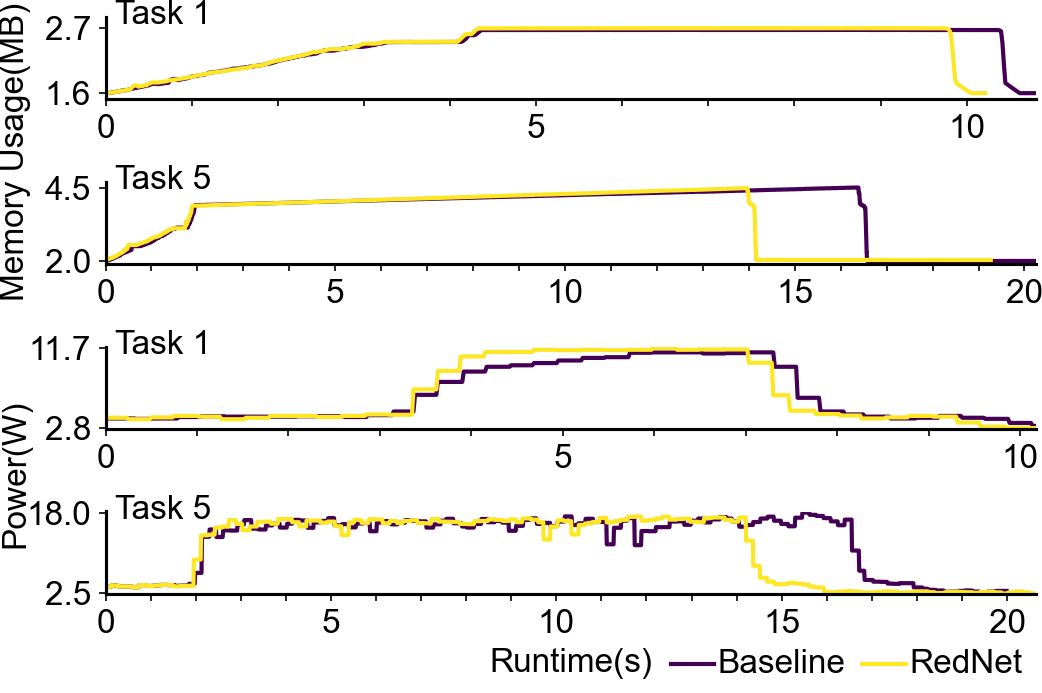}
    \vspace{-2ex}
    \caption{\name~can reduce power consumption with a tiny extra memory footprint for Task 1 and Task 5.}
    \label{fig:power56}
    \vspace{-3ex}
\end{figure}

Besides, we list the memory usage and power consumption of Task5 in \figurename~\ref{fig:power56} according to the payload logs. 
Here we choose a threshold of 0.3 (\tablename~\ref{tab:yolo_different_exit_threshold}) as an example. 
\name~can clearly reduce total energy consumption (i.e., the area under the power line) with almost no extra memory. 

\subsection{Hardware-in-the-loop Emulation}
\label{sec:eval:emulation}

{In this section, we use our hardware-in-the-loop radiation emulator to validate \name's tolerance to various space radiation-induced errors.
We first illustrate the setup information and metrics of evaluating the error-tolerance ability. 
Then, we list the results of tolerating errors with two cases, i.e., a certain number of bit errors randomly appear in run DNN model's memory space and a small number of bit errors occur intensively in the error-sensitive memory space.}

\subsubsection{{Evaluation Setup and Metrics}}
\label{subsubsec:metrics}

\noindent\textbf{Bit errors in global area v.s. the sensitive area of memory.} 
In our sensitivity analysis in~\S~\ref{subsec:senanaly}, we scan the whole memory space (i.e., \textit{global area}) of the runtime DNN model. 
We observe that the bit-error sensitivity is distributed unevenly, such as the shallow DNN layers (i.e., \textit{sensitive area}) are more sensitive to bit errors \S~\ref{subsec:unevensensitive}. 
Thus, we consider two cases for injecting bit errors to evaluate \name. 
First, we randomly inject a certain number of bit errors (with pre-defined error settings) to the global area for the runtime DNN model's memory space (\S~\ref{subsec:global}). 
This comes with the ground radiation tests~\cite{luo2021influence}, in which radiation flips bits in arbitrary locations in memory. 
Second, we consider a more challenging case in which a small number of bit errors (e.g., 5 bits) appear only in the sensitive area (\S~\ref{subsec:sensitivearea}). 
Please note this case could happen occasionally but is more challenging to deal with.

\noindent\textbf{Radiation-induced bit error settings.}
The bit-flip errors generated by our radiation emulator (\S~\ref{subsec:setup}) are conforming to the configurable \textit{error model} (\S~\ref{subsec: configurations}).
In the global memory area case, we have 3-bit error settings following~\cite{fabero2020single}, i.e. totally flipping 100, 200, and 500 bits. 
For each setting, we use this error model to inject bit errors considering single- and multi-bit errors.  
For example, when flipping 200 bits, there could be $16\times$ 2-bit errors, $4\times$ 3-bit errors, $1\times$ 6-bit errors, and $1\times$ 8-bit errors and $142\times$ 1-bit errors. 
After flipping bits with our hardware-in-the-loop emulator, we test \name's error tolerance ability. 
We run 500 times such tests for each task to avoid randomness (we show 100 of them due to limited space but the results are similar).
In the sensitivity memory area, we also use 3-bit error settings, i.e. flipping 5, 50, and 100 bits following the same error model. 
We use our emulator to perform this bit error injection step.

\noindent\textbf{Metrics.}
We use 2 metrics for \name's error-tolerance. 
First, we record how a DNN model's performance is influenced by bit errors. 
For Task 1-4, we calculate the accuracy (\%) drop on the test set and we calculate the mAP (\%)~\cite{everingham2010pascal} drop for Task 5. 
Since various orbital computing tasks are safety-sensitive (e.g., disaster monitoring), a more straightforward metric is needed to measure if a DNN model in orbit can be trustworthy or not. 
Thus, we use \textit{model crash} (defined as accuracy or mAP drop 10\%) \cite{hong2019terminal} to show how the above bit errors influence the decision-making of runtime DNN models.

\noindent\textbf{Baselines.} We consider the clean model without any protections (denoted as Clean) and one widely used error-tolerant method: \textit{clipping the output of ReLU activations} (denoted as Clip) used in~\cite{chen2021low, hoang2020ft} as our baselines.
We note that other existing mitigations ~\cite{ning2021ftt, li2020defending} targeted on dumb random errors cannot work well in runtime error bit-flips.

\subsubsection{Radiation Tolerance in Global Memory Areas}
\label{subsec:global} 

\begin{figure}[!htbp]
    \centering
    \includegraphics[width=0.88\linewidth]{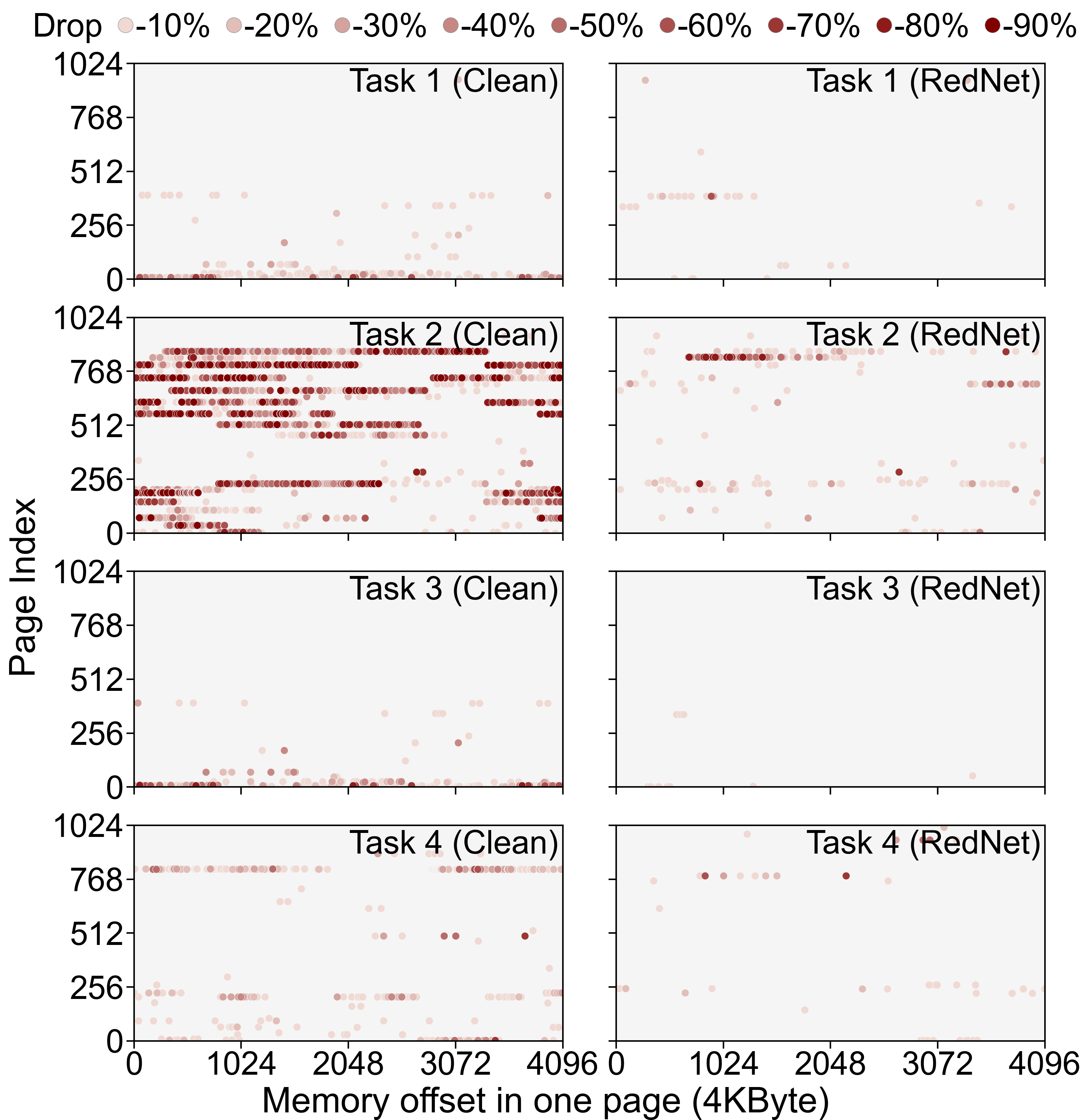} 
    \vspace{-2ex}
    \caption{Radiation sensitive bits of 5 tasks in their runtime DNN models' global memory area (Clean v.s. \name).}
    \label{fig:loop} \vspace{-3ex}
\end{figure}

First, recall the sensitivity analysis in \S~\ref{fig:task1loop}, we use our radiation emulator (\S~\ref{subsec:setup}) to comprehensively study the 5 tasks' sensitivity in global memory space (bits that are sensitive to errors) and report the results in \figurename~\ref{fig:loop}.  
We can observe that the sensitive bits are significantly reduced by using \name~which indicates its ability of radiation tolerance.

\begin{figure}[!htbp]
\vspace{-1ex}
    \centering
    \includegraphics[width=\linewidth]{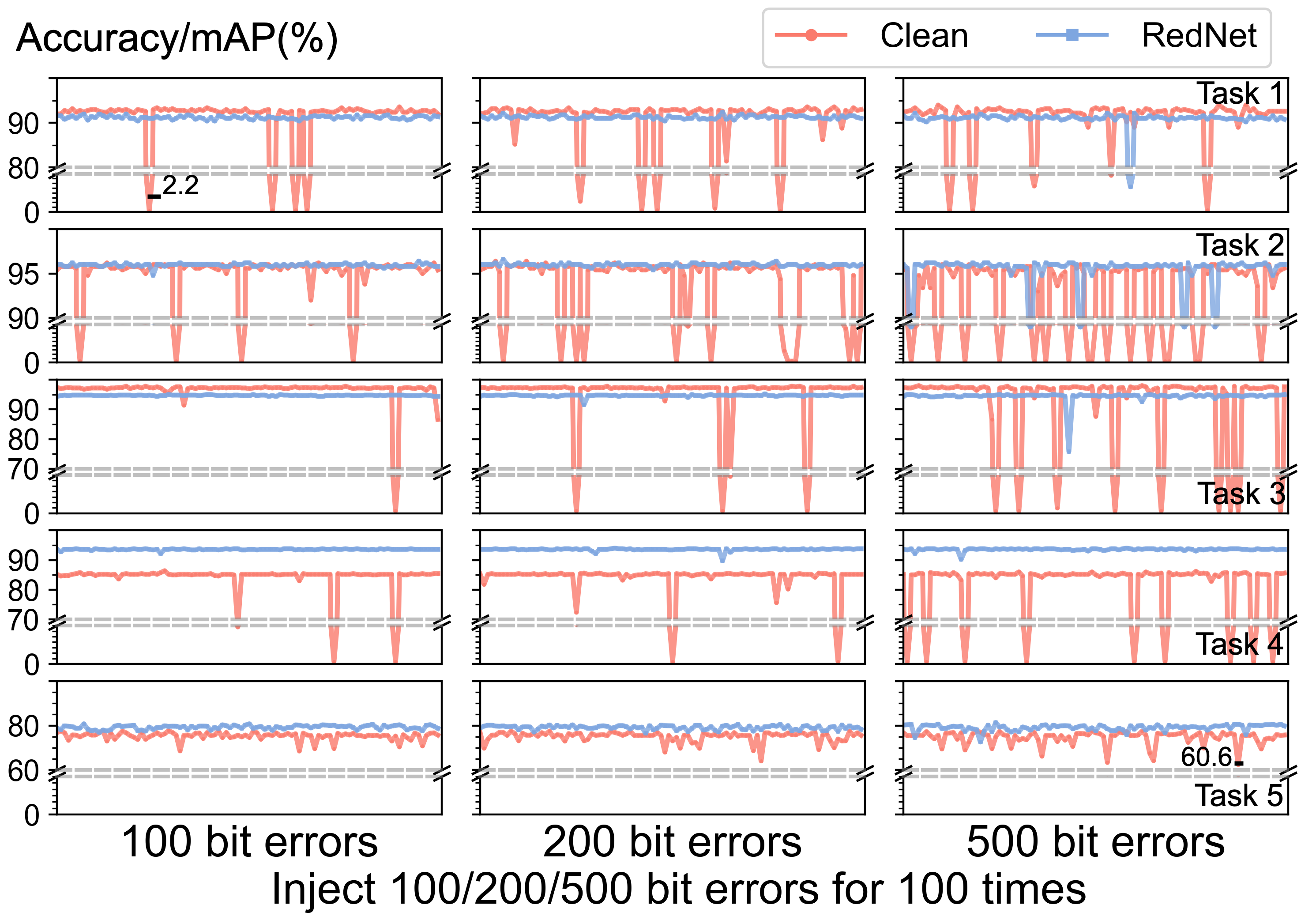}
    \vspace{-4ex}
    \caption{Radiation-induced bit error tolerance under 100, 200, and 500 error settings in the global memory area.}
    \label{fig:globaltest_a}
    \vspace{-2ex}
\end{figure}

In \figurename~\ref{fig:globaltest_a}, we display the accuracy/mAP of our 5 tasks (i.e., Y-axis) under 100, 200, and 500-bit errors in terms of test rounds (i.e., X-axis). 
We can observe that the DNN model without protection (i.e., Clean) clearly experiences a performance drop in all tasks. 
Particularly, the 100-bit error caused by emulated radiation can let the accuracy of Task 1 drop from 92\% to 2.2\%.
However, \name~can maintain the performance of all 5 tasks in almost all the tests. 

\begin{figure}[!htbp]
    \centering
    \includegraphics[width=\linewidth]{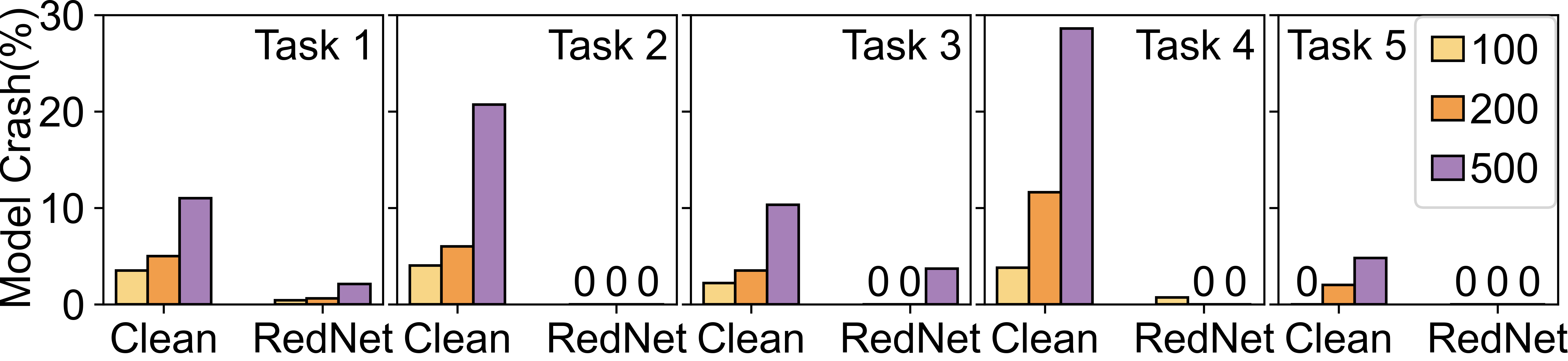}
    \vspace{-4ex}
    \caption{DNN model crash mitigation in the global area.}
    \label{fig:globaltest_b}
    \vspace{-3ex}
\end{figure}

We also list the results considering the model crash in \figurename~\ref{fig:globaltest_b}.  
The model crash in Clean reaches more than 20\% out of the 500 tests for Task 2 and Task 4, and more than 10\% in Task 1 and Task 3 under 500-bit flips.
We find that~\name~can suppress the model crash to 0.

We can observe that Task 5 is less sensitive to radiation-caused bit errors, which appears in a 4.8\% model crash and its mAP drops 17.7\% under 500-bit flips in Clean. 
However, Task 5 is used to detect objects which usually serves more safety-sensitive purposes. 
A drop of 17.7\% can lead to serious consequences, such as missing some important objects and wrong targeting (\figurename~\ref{fig:yoloerror}). 
We observe that \name~can maintain a better mAP compared with Clean (72.6\% v.s. 60.6\%) in the worst case of the 500 error bits test. 

\subsubsection{Radiation Tolerance in Sensitive Memory Areas}
\label{subsec:sensitivearea}

\begin{figure}[!htbp]
\vspace{-1ex}
    \centering
    \includegraphics[width=\linewidth]{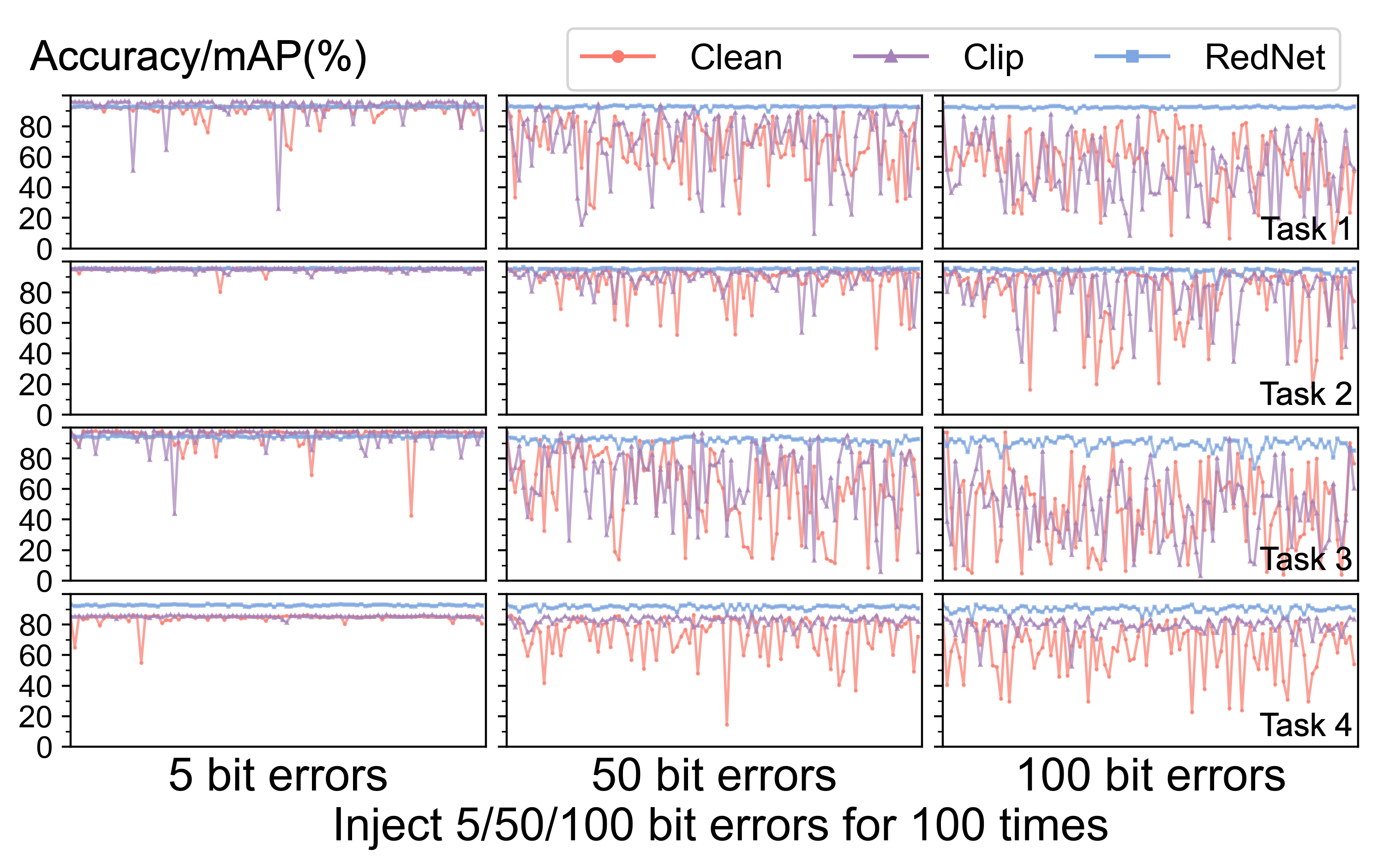}
    \vspace{-4ex}
    \caption{Radiation-induced bit error tolerance under 5, 50, and 100 error settings in the sensitive memory area.}
    \label{fig:sensitivetest_a}
    \vspace{-2ex}
\end{figure}

We also report the results in this more challenging case in which a small number of bit errors appear only in the sensitive memory area in \figurename~\ref{fig:sensitivetest_a}. 
Note that in each task, we choose sensitive memory areas according to \figurename~\ref{fig:loop}. 
For instance, for Task 1 and Task 3, we choose several consecutive pages starting from 13,000 bytes as sensitive areas. 
For Task 2 and Task 4, we choose several consecutive memory pages starting from 95,000 bytes as sensitive areas. 
As in \figurename~\ref{fig:sensitivetest_a}, errors in sensitive areas make a significantly more severe accuracy drop for Clean and Clip than in the global area case (\figurename~\ref{fig:globaltest_a}). 
We can observe that even very few bit errors occur in the sensitive area leading to a significant accuracy drop. 

\begin{figure}[!htbp]
    \centering
    \includegraphics[width=\linewidth]{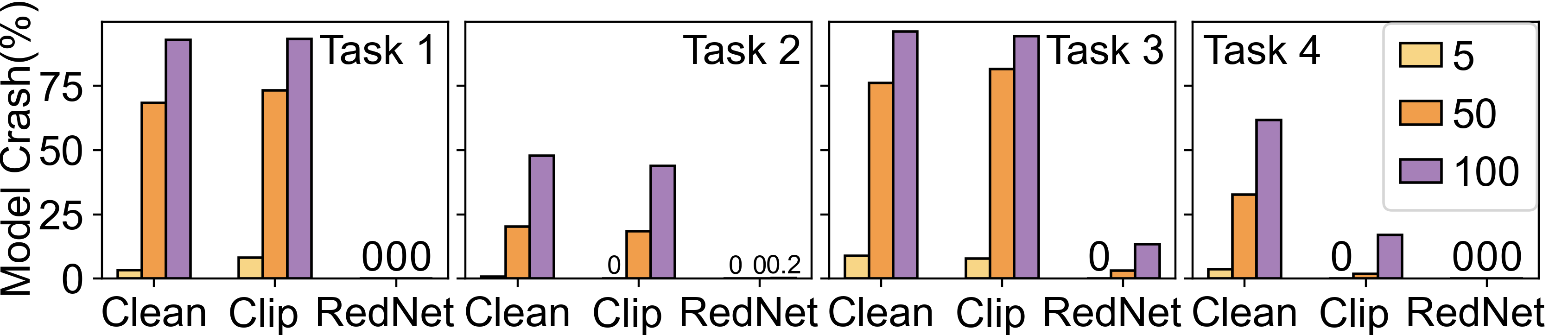}
   \vspace{-4ex}
    \caption{DNN model crash mitigation in the sensitive area.}
    \label{fig:sensitivetest_b}
    \vspace{-2ex}
\end{figure}

We also list the results of the model crash rate in \figurename~\ref{fig:sensitivetest_b} which indicates that bit errors in sensitive areas can lead to serious model crashes (e.g., 93.0\% crash with 100 bit errors in Task 1). 
We can also observe that existing model-layer protection without hardware-in-the-loop (e.g., Clip) cannot protect these tasks against radiation-caused bit errors. 
We further report the tolerance of ~\name~ under different memory configurations based on our emulator in Appendix~\ref{appen:exper}.

\section{Related Work}
\label{sec:related}

Orbital computing has attracted considerable attention recently~\cite{tao2024known,xing2024deciphering,wang2023mars,denby2023kodan,giuffrida2021varphi,denby2020orbital,zhang2024resource}. 
Due to the rapid increase of LEO satellite numbers and using COTS devices as components for satellites, COTS devices-enhance orbital computing has been a promising direction to save satellite communication bandwidth usage \cite{denby2023kodan,denby2020orbital} and accelerate latency-sensitive applications in-orbit (e.g. disaster monitoring)~\cite{tao2024known,zhang2024resource,giuffrida2021varphi}. 
Existing works can be classified into three aspects, including coordinating computation tasks between satellites or satellites with ground stations \cite{tao2024known,denby2023kodan,denby2020orbital}, measuring COTS computing device in-orbit~\cite{xing2024deciphering} (e.g. thermal control, power management, and performance), and accelerating applications like DNNs only on satellites~\cite{zhang2024resource,giuffrida2021varphi}. 
Instead, this paper studies an issue orthogonal to these works, namely, fault-tolerant in-orbit computing using COTS satellites in harsh space environments. 

In the context of fault-tolerant orbital computing, there have been some basic solutions from space radiation-hardened satellite hardware~\cite{wang2023mars,geist2023nasa,leppinen2017current} and system software protection~\cite{saramud2022implementation,wang2023mars} perspectives.
As demonstrated in \S\ref{sec:dnnanalysis}, these efforts are either expensive for COTS nanosatellites or overwhelming for upper-layer applications like DNN-based tasks.
To this end, our work explores a complementary {\em application-aware} radiation tolerance approach and demonstrates its effectiveness for COTS satellite-based DNN applications.
Besides, we notice other error-tolerance solutions for DNN models like the baseline in this paper~\cite{chen2021low, hoang2020ft} are not capable of DNN models in-orbit since they are not targeting solving the radiation-caused bit errors.

\section{Conclusion}
This paper explores reliable orbital computing using low-cost COTS satellite hardware when exposed to radiation hazards in harsh space environments.
We demonstrate that traditional general-purpose hardware and system software protections against space radiations are expensive and overwhelming for recent COTS nanosatellites due to their heavy reliance on resource redundancy and frequent service reboots.
Instead, we make a case for application-aware, cost-effective radiation tolerance for COTS satellite computing.
Our solution, \name, enhances prevalent DNN-based in-orbit tasks with application-aware radiation tolerance by exploiting their uneven sensitivity to bit errors and multi-bit errors' spatial correlation. 
We hope our lessons can foster more academic and industrial efforts on reliable orbital computing clusters.

\bibliographystyle{unsrt}
\bibliography{sample-base}

\begin{thebibliography}{10}

\bibitem{esposito2019highly}
M~Esposito, BC~Dominguez, M~Pastena, N~Vercruyssen, SS~Conticello, C~van Dijk, PF~Manzillo, and R~Koeleman.
\newblock Highly integration of hyperspectral, thermal and artificial intelligence for the esa phisat-1 mission.
\newblock In {\em Proceedings of the International Astronautical Congress IAC, Washington, DC, USA}, pages 21--25, 2019.

\bibitem{adams2019towards}
Caleb Adams, Allen Spain, Jackson Parker, Matthew Hevert, James Roach, and David Cotten.
\newblock Towards an integrated gpu accelerated soc as a flight computer for small satellites.
\newblock In {\em 2019 IEEE Aerospace Conference}, pages 1--7. IEEE, 2019.

\bibitem{starDetect}
{StarDetect}.
\newblock {GPU-enabled Payload, Website in Chinese}.
\newblock \url{http://www.stardetect.cn/h-col-127.html}, 2024.
\newblock Online; accessed 4-May-2024.

\bibitem{cappaert2018building}
Jeroen Cappaert.
\newblock Building, deploying and operating a cubesat constellation-exploring the less obvious reasons space is hard.
\newblock 2018.

\bibitem{denby2020orbital}
Bradley Denby and Brandon Lucia.
\newblock {Orbital Edge Computing: Nanosatellite Constellations as a New Class of Computer System}.
\newblock In {\em {Proceedings of the Twenty-Fifth International Conference on Architectural Support for Programming Languages and Operating Systems (ASPLOS)}}, pages 939--954, 2020.

\bibitem{denby2019orbital}
Bradley Denby and Brandon Lucia.
\newblock Orbital edge computing: Machine inference in space.
\newblock {\em IEEE Computer Architecture Letters}, 18(1):59--62, 2019.

\bibitem{denby2023kodan}
Bradley Denby, Krishna Chintalapudi, Ranveer Chandra, Brandon Lucia, and Shadi Noghabi.
\newblock Kodan: Addressing the computational bottleneck in space.
\newblock In {\em Proceedings of the 28th ACM International Conference on Architectural Support for Programming Languages and Operating Systems, Volume 3}, pages 392--403, 2023.

\bibitem{tao2024known}
Bill Tao, Om~Chabra, Ishani Janveja, Indranil Gupta, and Deepak Vasisht.
\newblock Known knowns and unknowns: Near-realtime earth observation via query bifurcation in serval.
\newblock In {\em 21st USENIX Symposium on Networked Systems Design and Implementation (NSDI 24)}, pages 809--824. USENIX Association, 2024.

\bibitem{leyva2023satellite}
Israel Leyva-Mayorga, Marc Martinez-Gost, Marco Moretti, Ana Pe{\'r}ez-Neira, Miguel~{\'A}ngel V{\'a}zquez, Petar Popovski, and Beatriz Soret.
\newblock Satellite edge computing for real-time and very-high resolution earth observation.
\newblock {\em IEEE Transactions on Communications}, 2023.

\bibitem{wiseman2019impact}
Matthew Wiseman and Alice Bradley.
\newblock Impact of the length of the sea ice-free summer season on alaskan arctic coastal erosion rates.
\newblock In {\em AGU Fall Meeting Abstracts}, volume 2019, pages C13D--1345, 2019.

\bibitem{kong2019monitoring}
Juwon Kong, Youngryel Ryu, Rasmus Houborg, and Minseok Kang.
\newblock Monitoring canopy photosynthesis in high spatial and temporal resolution using cubesat imagery.
\newblock In {\em AGU Fall Meeting Abstracts}, volume 2019, pages B11F--2397, 2019.

\bibitem{pri2024}
Private correspondence with leo smallsat operator, 2024.

\bibitem{wang2023mars}
Haoda Wang, Steven Myint, Vandi Verma, Yonatan Winetraub, Junfeng Yang, and Asaf Cidon.
\newblock {Mars Attacks! Software Protection Against Space Radiation}.
\newblock In {\em {Proceedings of the 22nd ACM Workshop on Hot Topics in Networks (HotNets)}}, pages 245--253, 2023.

\bibitem{xing2024deciphering}
Ruolin Xing, Mengwei Xu, Ao~Zhou, Qing Li, Yiran Zhang, Feng Qian, and Shangguang Wang.
\newblock {Deciphering the Enigma of Satellite Computing with COTS Devices: Measurement and Analysis}.
\newblock In {\em {ACM International Conference on Mobile Computing and Networking (MobiCom)}}, 2024.

\bibitem{choumos2024artificial}
George Choumos, Konstantinos Tsaprailis, Vaios Lappas, and Charalampos Kontoes.
\newblock Artificial intelligence for a safe space: Data and model development trends in orbit prediction and collision avoidance.
\newblock In {\em AIAA SCITECH 2024 Forum}, page 2066, 2024.

\bibitem{bezerra2011carmen2}
Fran{\c{c}}oise Bezerra, Robert Ecoffet, Eric Lorf{\`e}vre, Anne Samaras, and Christelle Deneau.
\newblock {CARMEN2/MEX}: An in-flight laboratory for the observation of radiation effects on electronic devices.
\newblock In {\em 2011 12th European Conference on Radiation and Its Effects on Components and Systems}, pages 607--614. IEEE, 2011.

\bibitem{samaras2011carmen}
Anne Samaras, Fran{\c{c}}oise Bezerra, Eric Lorfevre, and Robert Ecoffet.
\newblock {CARMEN}-2: In flight observation of non destructive single event phenomena on memories.
\newblock In {\em 2011 12th European Conference on Radiation and Its Effects on Components and Systems}, pages 839--848. IEEE, 2011.

\bibitem{walters2013practical}
John~Paul Walters, Kenneth~M Zick, and Matthew French.
\newblock A practical characterization of a {NASA} {S}pace{C}ube application through fault emulation and laser testing.
\newblock In {\em 2013 43rd Annual IEEE/IFIP International Conference on Dependable Systems and Networks (DSN)}, pages 1--8. IEEE, 2013.

\bibitem{fabero2020single}
Juan~Carlos Fabero, Hortensia Mecha, Francisco~J Franco, Juan~Antonio Clemente, Golnaz Korkian, Solenne Rey, Benjamin Cheymol, Maud Baylac, Guillaume Hubert, and Raoul Velazco.
\newblock Single event upsets under 14-mev neutrons in a 28-nm {SRAM}-based {FPGA} in static mode.
\newblock {\em IEEE Transactions on Nuclear Science}, 67(7):1461--1469, 2020.

\bibitem{su2021triple}
Haomiao Su, Tiejun Lu, Changlei Feng, and Lei Chen.
\newblock Triple module redundancy reliability framework design based on heterogeneous multi-core processor.
\newblock {\em Procedia Computer Science}, 183:504--511, 2021.

\bibitem{spaceXRocket}
{SpaceX}.
\newblock {Dragon's Radiation-Tolerant Design}.
\newblock \url{https://aviationweek.com/dragons-radiation-tolerant-design}, 2024.
\newblock Online; accessed 5-May-2024.

\bibitem{zhou2022carbink}
Yang Zhou, Hassan~MG Wassel, Sihang Liu, Jiaqi Gao, James Mickens, Minlan Yu, Chris Kennelly, Paul Turner, David~E Culler, Henry~M Levy, et~al.
\newblock Carbink:$\{$Fault-Tolerant$\}$ far memory.
\newblock In {\em 16th USENIX Symposium on Operating Systems Design and Implementation (OSDI 22)}, pages 55--71, 2022.

\bibitem{schmidt2017radiation}
Andrew~G Schmidt, Matthew French, and Thomas Flatley.
\newblock Radiation hardening by software techniques on fpgas: Flight experiment evaluation and results.
\newblock In {\em 2017 IEEE Aerospace Conference}, pages 1--8. IEEE, 2017.

\bibitem{skeggs2022vivid}
Cel~Andromeda Skeggs.
\newblock {\em Vivid: An Operating System Kernel for Radiation-Tolerant Flight Control Software}.
\newblock PhD thesis, Massachusetts Institute of Technology, 2022.

\bibitem{saramud2022implementation}
MV~Saramud, MV~Karaseva, IV~Kovalev, and VV~Losev.
\newblock Implementation of processor cores of a fault-tolerant control system on fpga under external control.
\newblock In {\em Journal of Physics: Conference Series}, volume 2388, page 012048. IOP Publishing, 2022.

\bibitem{Chaohu-1}
{Spacety}.
\newblock {Chaohu-1 SAR satellite}, 2022.

\bibitem{yue2017single}
Guo Yue, Wang Shaojun, Ma~Ning, Li~Pan, and Peng Yu.
\newblock {A Single Event Latch-up protection method for SRAM FPGA}.
\newblock In {\em 2017 13th IEEE International Conference on Electronic Measurement \& Instruments (ICEMI)}, pages 332--336. IEEE, 2017.

\bibitem{caron2019physical}
P~Caron, C~Inguimbert, L~Artola, R~Ecoffet, and F~Bezerra.
\newblock Physical mechanisms of proton-induced single-event upset in integrated memory devices.
\newblock {\em IEEE Transactions on Nuclear Science}, 66(7):1404--1409, 2019.

\bibitem{nidhin2017understanding}
TS~Nidhin, Anindya Bhattacharyya, RP~Behera, T~Jayanthi, and K~Velusamy.
\newblock Understanding radiation effects in sram-based field programmable gate arrays for implementing instrumentation and control systems of nuclear power plants.
\newblock {\em Nuclear Engineering and Technology}, 49(8):1589--1599, 2017.

\bibitem{bezerra202314}
Fran{\c{c}}oise Bezerra, Marine Ruffenach, Robert Ecoffet, J{\'e}r{\^o}me Carron, Julien Mekki, L{\'e}o Co{\"\i}c, and J{\'e}r{\'e}my Guillermin.
\newblock 14 years of in-flight experimental data on carmen-mex.
\newblock {\em IEEE Transactions on Nuclear Science}, 2023.

\bibitem{dutta2007multiple}
Avijit Dutta and Nur~A Touba.
\newblock Multiple bit upset tolerant memory using a selective cycle avoidance based sec-ded-daec code.
\newblock In {\em 25th IEEE VLSI Test Symposium (VTS'07)}, pages 349--354. IEEE, 2007.

\bibitem{zaharia2012resilient}
Matei Zaharia, Mosharaf Chowdhury, Tathagata Das, Ankur Dave, Justin Ma, Murphy McCauly, Michael~J Franklin, Scott Shenker, and Ion Stoica.
\newblock Resilient distributed datasets: A $\{$Fault-Tolerant$\}$ abstraction for $\{$In-Memory$\}$ cluster computing.
\newblock In {\em 9th USENIX symposium on networked systems design and implementation (NSDI 12)}, pages 15--28, 2012.

\bibitem{underwood1990orbit}
Craig Underwood.
\newblock In-orbit radiation effects monitoring on the uosat satellites.
\newblock {\em Technical Report}, 1990.

\bibitem{radaelli2005investigation}
Daniele Radaelli, Helmut Puchner, Skip Wong, and Sabbas Daniel.
\newblock Investigation of multi-bit upsets in a 150 nm technology {SRAM} device.
\newblock {\em IEEE Transactions on Nuclear Science}, 52(6):2433--2437, 2005.

\bibitem{kobayashi2020scaling}
Daisuke Kobayashi.
\newblock Scaling trends of digital single-event effects: A survey of seu and set parameters and comparison with transistor performance.
\newblock {\em IEEE Transactions on Nuclear Science}, 68(2):124--148, 2020.

\bibitem{aditya2018seu}
Kritika Aditya, Rohit Saini, Manoj Kumar, Ramendra Singh, and Abhisek Dixit.
\newblock Seu sensitivity of a 14-nm soi finfet edram cell under heavy-ion irradiation.
\newblock In {\em 2018 4th IEEE International Conference on Emerging Electronics (ICEE)}, pages 1--4. IEEE, 2018.

\bibitem{kohler2017analysis}
P~Kohler, Vincent Pouget, Fr{\'e}d{\'e}ric Wrobel, F~Saigne, Pierre-Xiao Wang, and M-C Vassal.
\newblock Analysis of single-event effects in ddr3 and ddr3l sdrams using laser testing and monte-carlo simulations.
\newblock {\em IEEE Transactions on Nuclear Science}, 65(1):262--268, 2017.

\bibitem{geist2023nasa}
Alessandro Geist, Gary Crum, Cody Brewer, Dennis Afanasev, Sebastian Sabogal, David Wilson, Justin Goodwill, James Marshall, Noah Perryman, Nick Franconi, et~al.
\newblock Nasa spacecube next-generation artificial-intelligence computing for stp-h9-scenic on iss.
\newblock 2023.

\bibitem{leppinen2017current}
Hannu Leppinen.
\newblock Current use of linux in spacecraft flight software.
\newblock {\em IEEE Aerospace and Electronic Systems Magazine}, 32(10):4--13, 2017.

\bibitem{burcin2002rad750}
L~Burcin.
\newblock Rad750 experience: The challenge of see hardening a high performance commercial processor.
\newblock In {\em Microelectronics Reliability \& Qualification Workshop (MRQW)}, 2002.

\bibitem{dopson2005softecc}
Dave Dopson.
\newblock {\em SoftECC: A system for software memory integrity checking}.
\newblock PhD thesis, Massachusetts Institute of Technology, 2005.

\bibitem{favalli2004annotated}
Michele Favalli.
\newblock Annotated bit flip fault model.
\newblock In {\em 19th IEEE International Symposium on Defect and Fault Tolerance in VLSI Systems, 2004. DFT 2004. Proceedings.}, pages 366--376. IEEE Computer Society, 2004.

\bibitem{wu2019protecting}
Xin-Chuan Wu, Timothy Sherwood, Frederic~T Chong, and Yanjing Li.
\newblock Protecting page tables from rowhammer attacks using monotonic pointers in dram true-cells.
\newblock In {\em Proceedings of the Twenty-Fourth International Conference on Architectural Support for Programming Languages and Operating Systems}, pages 645--657, 2019.

\bibitem{saileshwar2022randomized}
Gururaj Saileshwar, Bolin Wang, Moinuddin Qureshi, and Prashant~J Nair.
\newblock Randomized row-swap: mitigating row hammer by breaking spatial correlation between aggressor and victim rows.
\newblock In {\em Proceedings of the 27th ACM International Conference on Architectural Support for Programming Languages and Operating Systems}, pages 1056--1069, 2022.

\bibitem{gupta2019xbd}
Ritwik Gupta, Richard Hosfelt, Sandra Sajeev, Nirav Patel, Bryce Goodman, Jigar Doshi, Eric Heim, Howie Choset, and Matthew Gaston.
\newblock xbd: A dataset for assessing building damage from satellite imagery.
\newblock {\em arXiv preprint arXiv:1911.09296}, 2019.

\bibitem{kim2015ramulator}
Yoongu Kim, Weikun Yang, and Onur Mutlu.
\newblock Ramulator: A fast and extensible {DRAM} simulator.
\newblock {\em IEEE Computer architecture letters}, 15(1):45--49, 2015.

\bibitem{rosenfeld2011dramsim2}
Paul Rosenfeld, Elliott Cooper-Balis, and Bruce Jacob.
\newblock Dramsim2: A cycle accurate memory system simulator.
\newblock {\em IEEE computer architecture letters}, 10(1):16--19, 2011.

\bibitem{li2020dramsim3}
Shang Li, Zhiyuan Yang, Dhiraj Reddy, Ankur Srivastava, and Bruce Jacob.
\newblock Dramsim3: A cycle-accurate, thermal-capable dram simulator.
\newblock {\em IEEE Computer Architecture Letters}, 19(2):106--109, 2020.

\bibitem{cheng2017remote}
Gong Cheng, Junwei Han, and Xiaoqiang Lu.
\newblock Remote sensing image scene classification: Benchmark and state of the art.
\newblock {\em Proceedings of the IEEE}, 105(10):1865--1883, 2017.

\bibitem{xia2017aid}
Gui-Song Xia, Jingwen Hu, Fan Hu, Baoguang Shi, Xiang Bai, Yanfei Zhong, Liangpei Zhang, and Xiaoqiang Lu.
\newblock Aid: A benchmark data set for performance evaluation of aerial scene classification.
\newblock {\em IEEE Transactions on Geoscience and Remote Sensing}, 55(7):3965--3981, 2017.

\bibitem{xia2018dota}
Gui-Song Xia, Xiang Bai, Jian Ding, Zhen Zhu, Serge Belongie, Jiebo Luo, Mihai Datcu, Marcello Pelillo, and Liangpei Zhang.
\newblock Dota: A large-scale dataset for object detection in aerial images.
\newblock In {\em Proceedings of the IEEE conference on computer vision and pattern recognition}, pages 3974--3983, 2018.

\bibitem{esposito2019orbit}
Marco Esposito and A~Zuccaro Marchi.
\newblock In-orbit demonstration of the first hyperspectral imager for nanosatellites.
\newblock In {\em International Conference on Space Optics—ICSO 2018}, volume 11180, pages 760--770. SPIE, 2019.

\bibitem{wang2022empirical}
Di~Wang, Jing Zhang, Bo~Du, Gui-Song Xia, and Dacheng Tao.
\newblock An empirical study of remote sensing pretraining.
\newblock {\em IEEE Transactions on Geoscience and Remote Sensing}, 2022.

\bibitem{heiselberg2020remote}
Henning Heiselberg and Andrzej Stateczny.
\newblock Remote sensing in vessel detection and navigation, 2020.

\bibitem{ghazouani2019multi}
Fethi Ghazouani, Imed~Riadh Farah, and Basel Solaiman.
\newblock A multi-level semantic scene interpretation strategy for change interpretation in remote sensing imagery.
\newblock {\em IEEE Transactions on Geoscience and Remote Sensing}, 57(11):8775--8795, 2019.

\bibitem{he2016deep}
Kaiming He, Xiangyu Zhang, Shaoqing Ren, and Jian Sun.
\newblock Deep residual learning for image recognition.
\newblock In {\em Proceedings of the IEEE conference on computer vision and pattern recognition}, pages 770--778, 2016.

\bibitem{huang2017densely}
Gao Huang, Zhuang Liu, Laurens Van Der~Maaten, and Kilian~Q Weinberger.
\newblock Densely connected convolutional networks.
\newblock In {\em Proceedings of the IEEE conference on computer vision and pattern recognition}, pages 4700--4708, 2017.

\bibitem{yolov5}
Glenn Jocher.
\newblock Yolov5 by ultralytics, 2020.

\bibitem{rakin2019bit}
Adnan~Siraj Rakin, Zhezhi He, and Deliang Fan.
\newblock Bit-flip attack: Crushing neural network with progressive bit search.
\newblock In {\em Proceedings of the IEEE/CVF International Conference on Computer Vision}, pages 1211--1220, 2019.

\bibitem{wang2023aegis}
Jialai Wang, Ziyuan Zhang, Meiqi Wang, Han Qiu, Tianwei Zhang, Qi~Li, Zongpeng Li, Tao Wei, and Chao Zhang.
\newblock Aegis: Mitigating targeted bit-flip attacks against deep neural networks.
\newblock In {\em USENIX Security Symposium}, 2023.

\bibitem{li2020defending}
Jingtao Li, Adnan~Siraj Rakin, Yan Xiong, Liangliang Chang, Zhezhi He, Deliang Fan, and Chaitali Chakrabarti.
\newblock Defending bit-flip attack through dnn weight reconstruction.
\newblock In {\em 2020 57th ACM/IEEE Design Automation Conference (DAC)}, pages 1--6. IEEE, 2020.

\bibitem{chen2021low}
Zitao Chen, Guanpeng Li, and Karthik Pattabiraman.
\newblock A low-cost fault corrector for deep neural networks through range restriction.
\newblock In {\em 2021 51st Annual IEEE/IFIP International Conference on Dependable Systems and Networks (DSN)}, pages 1--13. IEEE, 2021.

\bibitem{hoang2020ft}
Le-Ha Hoang, Muhammad~Abdullah Hanif, and Muhammad Shafique.
\newblock Ft-clipact: Resilience analysis of deep neural networks and improving their fault tolerance using clipped activation.
\newblock In {\em 2020 Design, Automation \& Test in Europe Conference \& Exhibition (DATE)}, pages 1241--1246. IEEE, 2020.

\bibitem{ning2021ftt}
Xuefei Ning, Guangjun Ge, Wenshuo Li, Zhenhua Zhu, Yin Zheng, Xiaoming Chen, Zhen Gao, Yu~Wang, and Huazhong Yang.
\newblock Ftt-nas: Discovering fault-tolerant convolutional neural architecture.
\newblock {\em ACM Transactions on Design Automation of Electronic Systems (TODAES)}, 26(6):1--24, 2021.

\bibitem{agarap2018deep}
Abien~Fred Agarap.
\newblock Deep learning using rectified linear units (relu).
\newblock {\em arXiv preprint arXiv:1803.08375}, 2018.

\bibitem{drew2000computing}
John~H Drew, Andrew~G Glen, and Lawrence~M Leemis.
\newblock Computing the cumulative distribution function of the kolmogorov--smirnov statistic.
\newblock {\em Computational statistics \& data analysis}, 34(1):1--15, 2000.

\bibitem{redmon2018yolov3}
Joseph Redmon and Ali Farhadi.
\newblock Yolov3: An incremental improvement.
\newblock {\em arXiv preprint arXiv:1804.02767}, 2018.

\bibitem{lin2017feature}
Tsung-Yi Lin, Piotr Doll{\'a}r, Ross Girshick, Kaiming He, Bharath Hariharan, and Serge Belongie.
\newblock Feature pyramid networks for object detection.
\newblock In {\em Proceedings of the IEEE conference on computer vision and pattern recognition}, pages 2117--2125, 2017.

\bibitem{jetson_xavier_nx}
{NVIDIA}.
\newblock {Jetson Xavier NX Developer Kit}.
\newblock \url{https://developer.nvidia.com/embedded/jetson-xavier-nx-devkit}, 2024.
\newblock Online; accessed 14-March-2024.

\bibitem{tensorrt}
{NVIDIA}.
\newblock {TensorRT}.
\newblock \url{https://developer.nvidia.com/tensorrt}, 2024.
\newblock Online; accessed 14-March-2024.

\bibitem{luo2021influence}
Yinhong Luo, Fengqi Zhang, Wei Chen, Lili Ding, and Tan Wang.
\newblock The influence of ion track characteristics on single-event upsets and multiple-cell upsets in nanometer {SRAM}.
\newblock {\em IEEE Transactions on Nuclear Science}, 68(5):1111--1119, 2021.

\bibitem{everingham2010pascal}
Mark Everingham, Luc Van~Gool, Christopher~KI Williams, John Winn, and Andrew Zisserman.
\newblock The pascal visual object classes (voc) challenge.
\newblock {\em International journal of computer vision}, 88:303--338, 2010.

\bibitem{hong2019terminal}
Sanghyun Hong, Pietro Frigo, Yi{\u{g}}itcan Kaya, Cristiano Giuffrida, and Tudor Dumitraș.
\newblock Terminal brain damage: Exposing the graceless degradation in deep neural networks under hardware fault attacks.
\newblock In {\em 28th USENIX Security Symposium (USENIX Security 19)}, pages 497--514, 2019.

\bibitem{giuffrida2021varphi}
Gianluca Giuffrida, Luca Fanucci, Gabriele Meoni, Matej Bati{\v{c}}, L{\'e}onie Buckley, Aubrey Dunne, Chris Van~Dijk, Marco Esposito, John Hefele, Nathan Vercruyssen, et~al.
\newblock The $\phi$-sat-1 mission: The first on-board deep neural network demonstrator for satellite earth observation.
\newblock {\em IEEE Transactions on Geoscience and Remote Sensing}, 60:1--14, 2021.

\bibitem{zhang2024resource}
Qiyang Zhang, Xin Yuan, Ruolin Xing, Yiran Zhang, Zimu Zheng, Xiao Ma, Mengwei Xu, Schahram Dustdar, and Shangguang Wang.
\newblock Resource-efficient in-orbit detection of earth objects.
\newblock In {\em IEEE INFOCOM 2023-IEEE Conference on Computer Communications}. IEEE, 2024.

\bibitem{jedec2014jedec}
JEDEC Solid State~Technology Association et~al.
\newblock {JEDEC} standard: Low power double data rate 4 ({LPDDR4}).
\newblock {\em JEDEC Standard JESD209-4}, 2014.

\bibitem{koppula2019eden}
Skanda Koppula, Lois Orosa, A~Giray Ya{\u{g}}l{\i}k{\c{c}}{\i}, Roknoddin Azizi, Taha Shahroodi, Konstantinos Kanellopoulos, and Onur Mutlu.
\newblock Eden: Enabling energy-efficient, high-performance deep neural network inference using approximate dram.
\newblock In {\em Proceedings of the 52nd Annual IEEE/ACM International Symposium on Microarchitecture}, pages 166--181, 2019.

\bibitem{makihara2000analysis}
A~Makihara, H~Shindou, N~Nemoto, S~Kuboyama, S~Matsuda, T~Oshima, T~Hirao, H~Itoh, S~Buchner, and AB~Campbell.
\newblock Analysis of single-ion multiple-bit upset in high-density drams.
\newblock {\em IEEE Transactions on Nuclear Science}, 47(6):2400--2404, 2000.

\bibitem{pessl2016drama}
Peter Pessl, Daniel Gruss, Cl{\'e}mentine Maurice, Michael Schwarz, and Stefan Mangard.
\newblock $\{$DRAMA$\}$: Exploiting $\{$DRAM$\}$ addressing for $\{$Cross-CPU$\}$ attacks.
\newblock In {\em 25th USENIX security symposium (USENIX security 16)}, pages 565--581, 2016.

\bibitem{kingma2014adam}
Diederik~P Kingma and Jimmy Ba.
\newblock Adam: A method for stochastic optimization.
\newblock {\em arXiv preprint arXiv:1412.6980}, 2014.

\bibitem{yang2020arbitrary}
Xue Yang and Junchi Yan.
\newblock Arbitrary-oriented object detection with circular smooth label.
\newblock In {\em Computer Vision--ECCV 2020: 16th European Conference, Glasgow, UK, August 23--28, 2020, Proceedings, Part VIII 16}, pages 677--694. Springer, 2020.

\bibitem{cuda}
{NVIDIA}.
\newblock {CUDA runtime API}.
\newblock \url{https://docs.nvidia.com/cuda/cuda-runtime-api/index.html}, 2024.

\end{thebibliography}

\clearpage
\appendix
\section{DRAM hierarchy and radiation influence}
\label{appen:dram}

Radiation effects in harsh outer space can lead to transient damage to DRAM-based memory in COTS devices, causing single-event upset (SEUs) and multiple-cell upsets (MCUs). We introduce the detailed hierarchy of DRAM to better illustrate the radiation-induced errors from the hardware layers to the applications layers across multiple memory layers.

\begin{figure}[h]
    \centering
    \includegraphics[width=\linewidth]{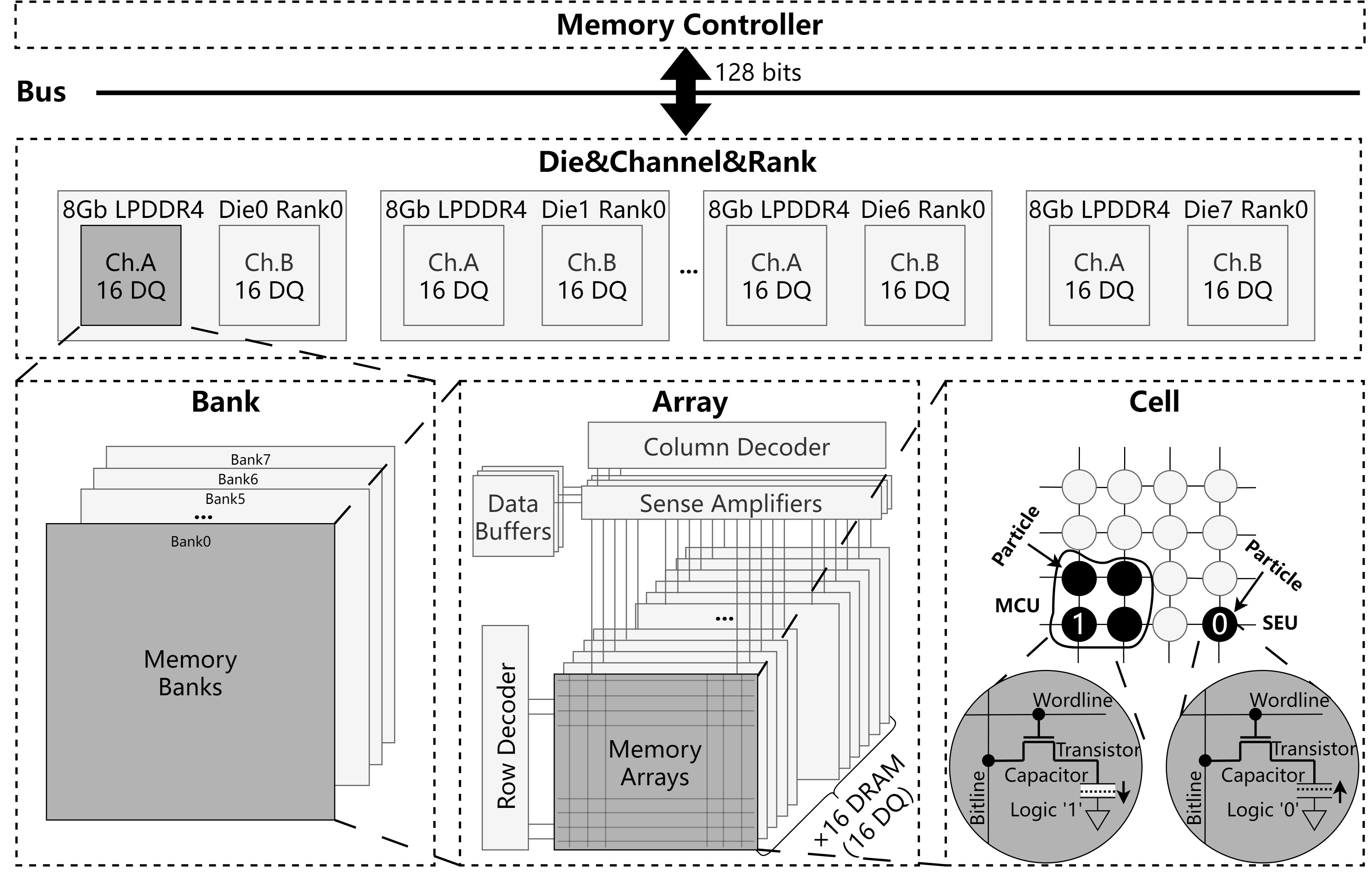}
    \caption{DRAM hierarchy and SEU/MCU influence.}
    \label{fig:dram hierarchy}
\end{figure}

\noindent\textbf{Dynamic random access memory (DRAM)} is widely used in COTS devices' main memory. 
We consider our computing payload, Jetson Xavier NX series~\cite{jetson_xavier_nx}, which uses 8GB 128-bit LPDDR4 (JESD209-4~\cite{jedec2014jedec}) as its main memory. 
DRAM typically consists of six hierarchies, including die, channel, rank, bank, array, and cell as illustrated in \figurename~\ref{fig:dram hierarchy}.
The memory density and the details of each hierarchy are different. For instance, LPDDR4~\cite{jedec2014jedec} defines multiple dual-channel dies of different memory densities from 4Gb to 32Gb to suit different memory sizes.
To expand the channel width of memory, multiple dual-channel dies can be used (e.g. 128-bit LPDDR4 uses 8 dual-channel dies as shown in \figurename~\ref{fig:dram hierarchy}).
Each DARM channel consists of multiple banks and each bank contains multiple 2D arrays to meet the width of DQ data bus (i.e., column width).
DRAM cells are organized in arrays and the cells in one column share a single bitline. When the bitline is activated, the transition will be turned on, and then the capacitor charges to store logic 1 or discharges to store logic 0~\cite{koppula2019eden}.

\begin{figure*}[h]
    \centering
    \includegraphics[width=0.7\textwidth]{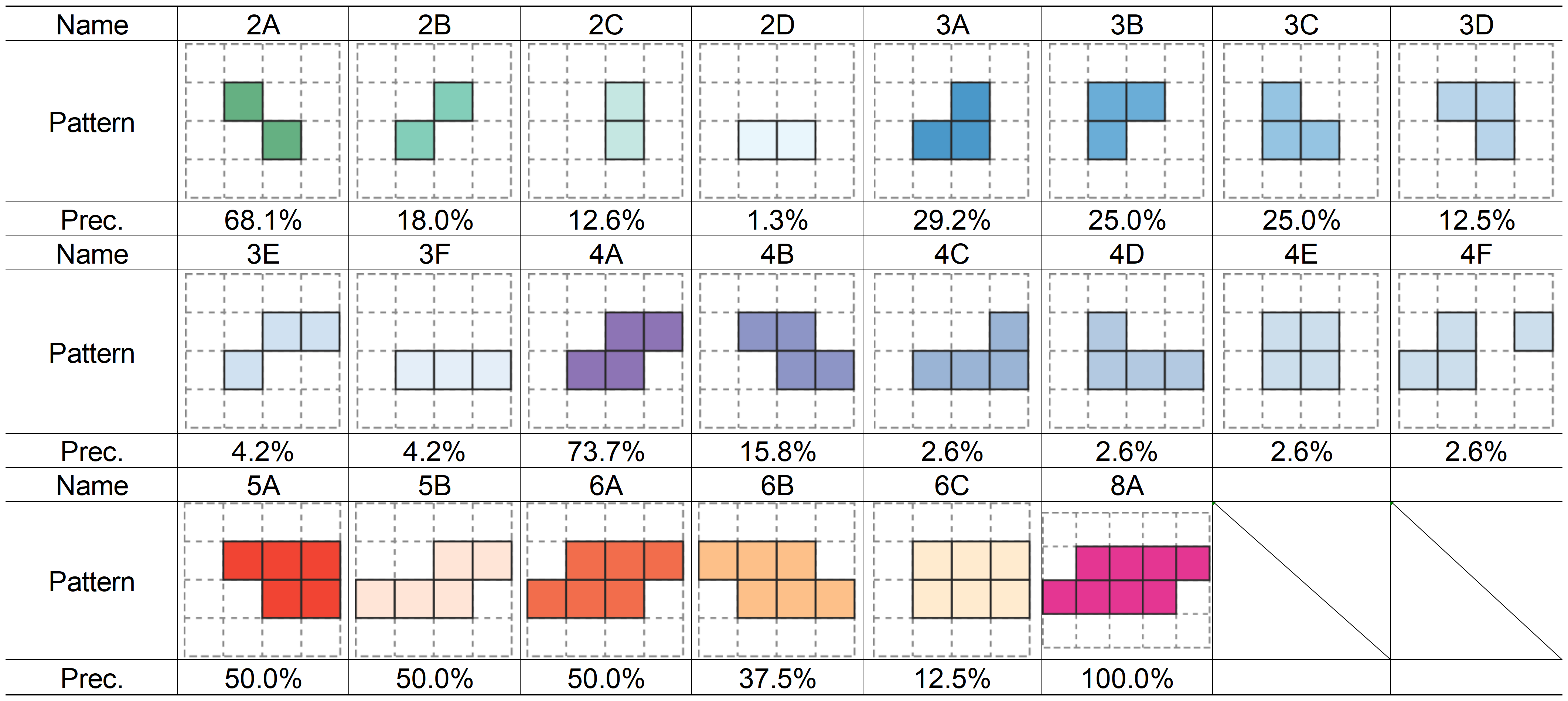}
    \caption{Different error patterns of multi-bit errors~\cite{fabero2020single}.}
    \label{fig:MCUs}
\end{figure*}

\noindent\textbf{Spatial contiguity across multiple memory layers.}
Multi-bit errors happen more frequently due to the advanced nanotechnology and more compact size of the chip.
Many studies have shown that many radiation-induced transient bit-flip errors manifest spatial contiguity in adjacent storage units (i.e., memory cells).
\figurename~\ref{fig:MCUs} shows a summary of various multiplicity (from 2 to 8) of MCUs collected in ~\cite{fabero2020single} under 14.2-MeV neutrons.
The multiple events tend to occur as pairs of cells in a diagonal orientation or along the wordline (i.e., two adjacent columns)~\cite{fabero2020single}.
Also, due to bit interleaving in modern memory systems, the adjacent cells/bits are not located in logically adjacent data bytes or data words~\cite{kohler2017analysis, bezerra2011carmen2}.
For example,~\cite{bezerra2011carmen2} recorded a 2-bit MCU flight data, which both flipped from the original byte "5A" to byte "5B" in two physical/logical addresses "10EC36" and "10EE36".
The two almost identical but not adjacent addresses are two adjacent cells struck by one particle.
Thus, the complicated error pattern of MCUs needs to be modeled across multi-layers as specified in \figurename~\ref{fig:dram hierarchy}.

\section{Design of Space Radiation Emulator}
\label{appen:remu}
\begin{figure}[h]
    \centering
    \includegraphics[width=0.98\linewidth]{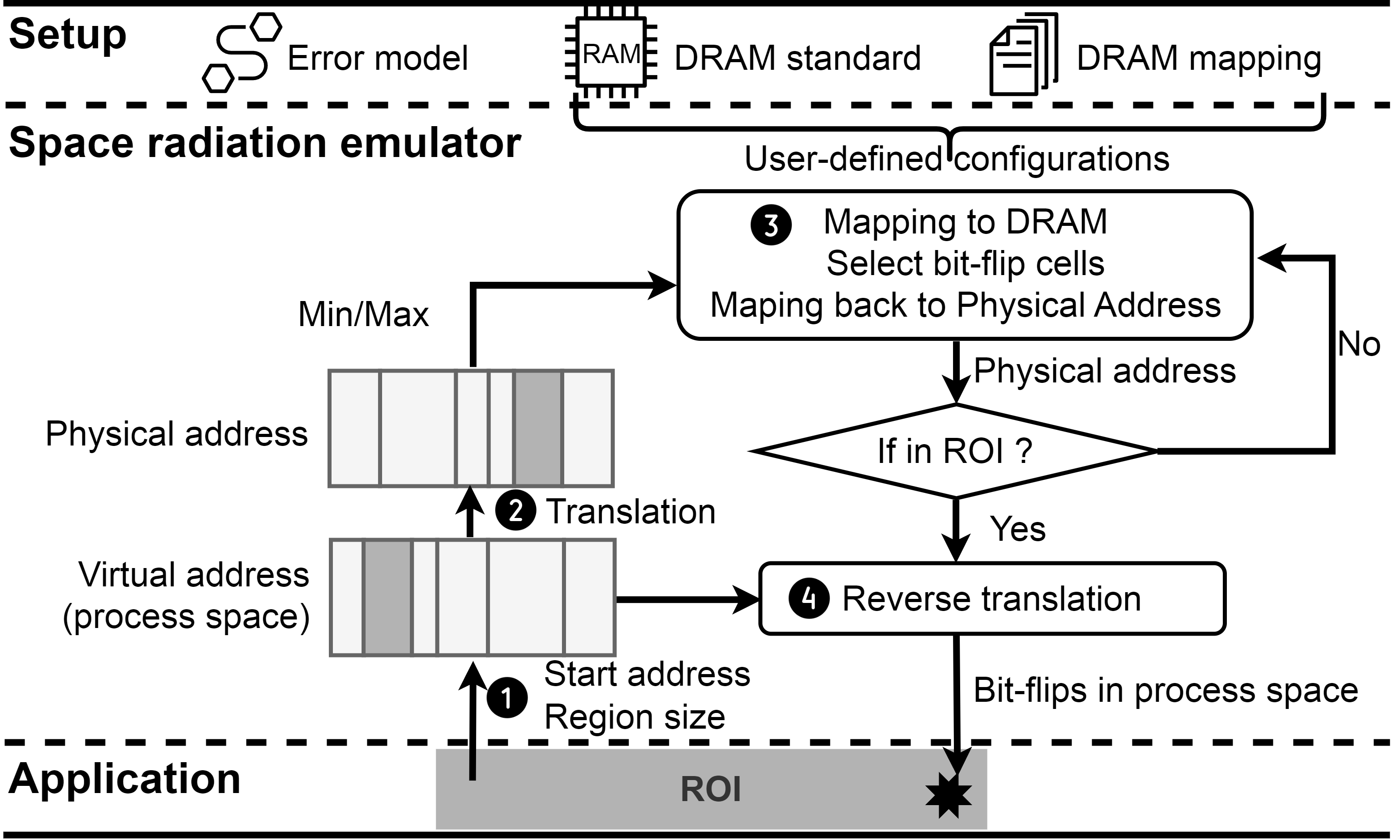}
    \caption{Details of space radiation error emulator.}
    \label{fig:overview}
\end{figure}

\figurename~\ref{fig:overview} illustrates the detailed workflow of our hardware-in-the-loop space radiation emulator. 
The inputs are user-defined, including the error model, DRAM standard, and DRAM mapping (\S\ref{subsec: configurations}).
The internal implementation has four critical parts.
First (\ding{182}), defining regions of interest (ROI) in the run-time program by setting the start address and the region size. 
Second (\ding{183}), making address translation from the process's virtual space to the physical address.
Third (\ding{184}), mapping the physical address to DRAM physical layouts (e.g. DRAM rows, columns, etc.), selecting bit-flip cells, and then mapping back to a set of physical addresses.
Fourth (\ding{185}), converting the set of physical addresses into a set of virtual addresses with the same offset in the corresponding block/page/frame.
Finally, flipping one bit of a byte at each address in the set of virtual addresses.
Note that to be more targeted, only the ROI of the program is tested under errors, thus ensuring the flips located in ROI is necessary.
The entire emulator is loaded as a dynamic-link library (\textit{libREMU\_mem.so}) which can be instrumented in any program.
In the following subsections, we give more details. 

\subsection{Extensible Configurations}\label{subsec: configurations}
\textbf{Error model.}
We give the \textit{error model} interface to statistically configure radiation-caused single- and multi-bit errors. 
According to the extensive ground radiation tests on memories, radiation-induced bit-flip errors differ in sizes, shapes, and frequencies~\cite{makihara2000analysis,fabero2020single,samaras2011carmen}. 
For single-bit errors, one particle striking results in a single bit-flip error. 
Single-bit errors are spatially independent from each other and can be modeled as random bit-flips.
For multi-bit errors, a particle may influence multiple adjacent cells resulting in a physical spatial adjacency of errors (i.e. adjacent rows and columns of cells in DRAM) with different sizes or shapes.
Following various previous statistics of ground radiation tests~\cite{fabero2020single}, we set the multi-bit error size ranging from 2 bits to 8 bits, and the probabilities of flipping two cells along one wordline is higher (80\%) than that along one bitline (20\%). 
Therefore, given a pair of reference row and column, only one or two adjacent rows but up to 5 adjacent columns can be affected. 
Such directionality is more likely to occur in multi-bit errors as shown in \figurename~\ref{fig:MCUs}.
Besides, single- and multi-bit errors will occur at the same time, in which single-bit errors have the highest probability of occurrence, followed by 2-bit errors and more than 6 bits are rare~\cite{fabero2020single}. 
Thus, following~\cite{fabero2020single}, we set the maximum frequency of 2-bit errors to be 12\%, 3-bit errors to be 2\%,  more than 3 bits to be 1\% and the rest are single-bit flips. 
Here we set our error model following real-world ground radiation tests~\cite{fabero2020single} but the error model in our emulator is configurable to adopt other radiation tests. 

\begin{figure*}[t]
    \centering
    \includegraphics[width=0.98\textwidth]{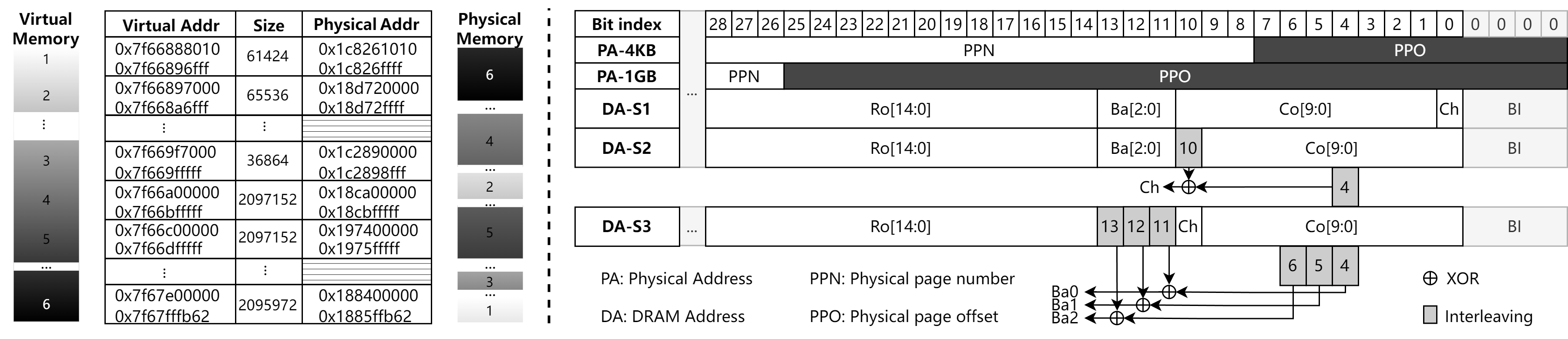}
    \caption{
    Left: page translation and pairs of virtual blocks and physical blocks. 
    Continuous virtual addresses both in and out of the virtual block, while continuous physical addresses in the physical block, and discontinuous physical addresses outside the physical block.
    Right: three addressing schemes: S1 is the default addressing scheme without addressing functions (i.e. XORs), while S2 and S3 consider the interleaving mode for bank and channel respectively~\cite{pessl2016drama}. 
    We set the lowest bit in addressing as $a_{4}$ because the burst length of LPDDR4 is 16 (i.e. BL16), in which continuously read and write 16-column data blocks to avoid row conflict, thus the size of the data block is $16\times W_{DQ}$ and is $16\times W_{CH}$ in one channel. 
    }
    \label{fig:mapping}
\end{figure*}

\noindent\textbf{DRAM configurations.}
There are many memory standards like DDR3, DDR4, LPDDR4, etc., with different specifications from different manufacturers (e.g. different memory densities per die). 
Moreover, the memory controller controls the address translation to get access to DRAM, but some manufacturers like Intel, do not disclose the algorithms.
Previous works reverse-engineer the DRAM addressing schemes for some standards like DDR3~\cite{pessl2016drama}. 
However, the existing results of reverse engineering cannot cover all the devices.
Therefore, our emulator uses \textit{DRAM standard} and \textit{DRAM mapping} as initial configurations for users to customize according to their own memory and knowledge. 
Following Ramulator~\cite{kim2015ramulator}, we set the numbers of valid channels, ranks, banks, rows, and columns to establish the DRAM hierarchy in \textit{DRAM standard}. 
Besides, \textit{DRAM mapping} defines the mapping from the DRAM hierarchy to the physical address (i.e., which bits in the address are the row index and column index). 
In our paper, expanding based on~\cite{kim2015ramulator}, we provide DRAM mappings with three addressing schemes (i.e. S1, S2, and S3) by considering LPDDR4~\cite{jedec2014jedec} flexibly. 
Details of these addressing schemes are in \figurename~\ref{fig:mapping} (right). 

\noindent\textbf{Source code instrumentation.}
The use of our emulator is plug-and-play through compiler-based instrumentation linking with the dynamic-link library \textit{libREMU\_mem.so}.
We expose a rich API for ease of use, and just two lines of code can achieve error injection into the ROI of process space.
Note that any program can be investigated by our emulator for error behaviors.
In this paper, we focus on the DNN inference program in TensorRT runtime~\cite{tensorrt}, which is commonly used in orbit for high-performance DNN inference.

\subsection{End-to-End Error Mapping}
\label{appen: mapping}
\textbf{From virtual blocks to physical blocks.}
After determining the ROI in process space, we use \textit{/proc/pid/pagemap\footnote{\url{https://www.kernel.org/doc/Documentation/vm/pagemap.txt}}} interface in the kernel to let a userspace process find out which
physical frame each virtual page is mapped to.
Note that in modern memory management, such as \textit{buddy system} in Linux, allocations are separate due to the limited physical space left when the process is running.
Thus, when allocated, a large virtually contiguous area, such as a DNN inference data buffer (i.e. our ROI), cannot be allocated to a large contiguous physical area unless the large page mechanism is enabled (normally disabled).
So the translating is necessary for run-time mapping following the kernel page table.

We denote the start address of ROI as $V_0$ and the size of ROI as $S_{ROI}$.
First, we get page size at runtime through standard C library \textit{sysconf\footnote{\url{https://man7.org/linux/man-pages/man3/sysconf.3.html}}}, usually 4KB.
Then, we find the corresponding physical frame number in the \textit{pagemap} based on the offset of $V_0$.
We should note that each frame number is page-aligned (i.e., frame size is equal to page size).
Then, from the beginning of $V_0$, for each page size in $S_{ROI}$, we get the corresponding physical frame number as above and combine the consecutive physical frames into one physical block.
Thus, in a pair of physical block and virtual block, either the physical block or the virtual block is an integer multiple of the page size, and either the physical addresses in the physical block or the virtual addresses in the virtual block are continuous.
In summary, we construct a one-to-one mapping of $M$ physical and $M$ virtual addresses/blocks ($\{P_i: V_i\}, i\in \{0,M-1\}$) in ROI as shown in \figurename~\ref{fig:mapping} (left).

\noindent\textbf{From physical addresses to DRAM.}
Based on the configurations of \textit{error model}, \textit{DRAM standard}, and \textit{DRAM mapping}, along with the converted physical addresses of ROI, we achieve an end-to-end mapping from DRAM bit-flip cells to process space.
We illustrate the high-level design of this mapping method with an example of 8GB 128-bit LPDDR4~\cite{jedec2014jedec}. 
The mapping is divided into three steps.

First, based on the maximum and minimum physical addresses of ROI, the range of each level involved in the DRAM hierarchy should be determined.
For example, as shown in \figurename~\ref{fig:mapping} (right), if the minimum physical address is 0 and the maximum is 4KB, only the lowest 12 bits of the physical address serve for addressing and it contains $2^0$ row and $2^7$ columns for S1, while $2^0$ row, $2^8$ columns, and $2^3$ banks for S3, in which the bank index is computed by XORing $a_{11}...a_{16}$ with the lower bits $a_{4}...a_{6}$ (i.e., column index).
This can parallel access memory by bank interleaving~\cite{pessl2016drama}. 

Second, we randomly select one row and one column in the range for error reference denoted as $r_0$ and $c_0$. 
With $r_0$ and $c_0$ as the origin, we get a set of bit-flips $\mathbf{B}=\{(r_i, c_i)\}_N, i\in \{0,N-1\}$ based on the error model, in which $N$ denotes the size of SEUs (i.e., $N=1$) or MCUs (i.e., $N>=2$). 
In the example above, only one row can be addressed, thus $r_0=-1$ and $c_0 \in \mathbb{Z} \cap [0, 2^7)$.
Besides, adjacency is represented by $r_i\pm 1$ and $c_i\pm1$.

Third, bit-flips in $\mathbf{B}$ are mapped back to their corresponding physical addresses. 
Because only row and column indexes $(r_i, c_i)$ can be determined in error models, while channels, ranks, and banks are lacking.
Thus, due to the principle of adjacency, we assume that the multiple bits in MCUs are in the same channel, rank, and bank (\figurename~\ref{fig:dram hierarchy}).
The interleaving mode should be considered when calculating the bank and the channel index.
For example, column indexes $a_{4}...a_{6}$ also serve as bank indexes by XOR in S3.
Finally, each bit-flip can be uniquely determined as a 5-dimension index, denoted as $(r_i, c_i, ch_i, ra_i, b_i)$ and can be converted into the corresponding physical address $PE_i$ as shown in \figurename~\ref{fig:mapping} (right).

Finally, the physical addresses $\{PE_i\}, i\in \{0,N-1\}$ are converted back to the virtual addresses $\{VE_i\}, i\in \{0,N-1\}$ in process space based on the mapping $\{P_i: V_i\}, i\in \{0,M-1\}$.
Specifically, first finding which physical block $PE_i$ is located in, then calculating the offset as $PE_i-P_i$, and $VE_i =  V_i + (PE_i-P_i)$.
As a result, all the addresses in $\{VE_i\}, i\in \{0,N-1\}$ are the targets to be flipped and the program will go on after flipping these bits.

\noindent\textbf{Region of interest (ROI).}
Because the entire process space in memory is too large, we set the region of interest (ROI) first.
As long as any address in $\{VE_i\}, i\in \{0, N-1\}$ is not in ROI, we will repeat the above DRAM mapping until all of them fall in ROI.
ROI can make the study of error behaviors more focused, which improves the efficiency of error emulation.
In the DNN inference program, our ROI is the TensorRT engine loaded in memory during the deserialization process, which occupies most of the time and memory during the entire inference.
Due to SEUs and MCUs being probabilistic events, the longer the time and the more memory footprint, the more likely they will happen. 

\section{Training Settings of \name}
\label{appen:training}
For training, we use the PyTorch platform on NVIDIA GeForce RTX 3080Ti GPUs. 
For scene recognition tasks (i.e., Task 1-4), the backbones are trained for no more than 100 epochs from the pre-trained models, and ICs are trained individually for another 200 epochs with an initial learning rate of 0.001 and Adam optimizer~\cite{kingma2014adam}, totally consuming 32 GPU hours.
For the object detection task (i.e., Task 5), we enhanced the capability to detect objects of arbitrary orientations by integrating angle classification and encoding methodologies from~\cite{yang2020arbitrary}~into the YOLOv5 architecture.
We split the DOTA dataset into 1024×1024 sub-images with a patch overlap of 200 pixels. 
Following the original YOLOv5 implementation~\cite{yolov5}, we pre-train the modified YOLOv5 model on the DOTA dataset for 200 epochs using an Adam optimizer with an initial learning rate of 0.01. 
We then attach the ICs to the model and fine-tune them for another 200 epochs, consuming 40 GPU hours in total.

\section{Speed-up settings of \name.} 
\label{appen:speed}
To meet the requirement of real-time processing in orbit. The acceleration of~\name~is necessary.
Leveraging TensorRT~\cite{tensorrt} and CUDA~\cite{cuda} APIs, we achieve the acceleration for both the activation functions and multi-exit strategy in ~\name.

First, the replacement of $\mathrm{LogClip}$ mentioned in \S~\ref{subsec: log} has no negative influence on inference speed. 
Because the activation is customized and implemented by TensorRT's plugin base class, \textit{IPluginV2IOExt}~\cite{tensorrt} with high performance.
Specifically, $\mathrm{LogClip}$ is implemented in CUDA multi-threads.

Second, to ensure the correct early exit sequence in the execution of \name, the TensorRT runtime inference should be further adjusted and optimized by the CUDA runtime API~\cite{cuda}, leveraging parallel stream and synchronization events.
Specifically, before each exit, we customize an \textit{exit layer} to record the synchronous event in the main CUDA stream, which indicates that the inference of the backbone has reached the exit point.
Therefore, the IC's/ID's inference can start from the exit point in a separate CUDA stream to parallel with the main CUDA stream.
This novel design first implements real multi-exit inference, bringing significant efficiency improvements.
We open source the design for ease of use (\url{https://anonymous.4open.science/r/RedNet-9665}).

\begin{table}[h]
\centering
\begin{tabular}{c|c|c|c}
\hline
\textbf{Mapping} & \textbf{N=5} & \textbf{N=50} & \textbf{N=100} \\ \hline
\textbf{S1} & 100 & 100 & 20 \\ \hline
\textbf{S2} & 100 & 73.4 & 45.3 \\ \hline
\textbf{S3} & 100 & 77.8 & 45.2 \\ \hline
\end{tabular}
\caption{Valid errors in sensitive areas of different memory configurations (S1, S2, and S3)}
\label{tab:invalid}
\end{table}

\section{Radiation tolerance of different memory configurations}
\label{appen:exper}
Based on our configurable emulator, we conduct extensive experiments on different configurations S1, S2, and S3 of LPDDR4~\cite{jedec2014jedec} as illustrated in Appendix~\ref{appen: mapping}.
As shown in Table~\ref{tab:invalid}, we find that S3 and S2 will be more likely to strike outside the sensitive area.
The valid (\%) indicates how many adjacent bits from the hardware layer are actually located in sensitive areas from process space.
When striking 100 bits 1000 times,  S1 is 100\% 100-bit errors located in the sensitive area, however, S2 is only 45.3\% valid 100-bit and S3 is only 45.2\% valid 100-bit in the sensitive area.
Interleaving is introduced in S2 and S3, which makes the adjacent columns located in separate banks and channels as shown in \figurename~\ref{fig:mapping} (right).
However, although the error behaviors change in S2 and S3, the valid bits are not substantially reduced, at least 46-bit in S2 and 45-bit in S3.
Thus, the model crash is not reduced and the accuracy is not improved because only a one-bit flip in the sensitive area is destructive.
We list the results of radiation tolerance of different memory configurations in Table~\ref{tab:exp}.

\begin{table*}[]
\centering
\resizebox{0.98\textwidth}{!}{
\begin{tabular}{c|c|c|cc|cc|cc|cc}
\hline
\multirow{2}{*}{\textbf{Mapping}} & \multirow{2}{*}{\textbf{Task}} & \multirow{2}{*}{\textbf{N}} & \multicolumn{2}{c|}{\textbf{Model   Crash(\%)}} & \multicolumn{2}{c|}{\textbf{Average   Accuracy(\%)}} & \multicolumn{2}{c|}{\textbf{Minimum   Accuracy(\%)}} & \multicolumn{2}{c}{\textbf{Maximum   Accuracy(\%)}} \\
 &  &  & \textbf{Baseline} & \textbf{RedNet} & \textbf{Baseline} & \textbf{RedNet} & \textbf{Baseline} & \textbf{RedNet} & \textbf{Baseline} & \textbf{RedNet} \\ \hline
\multirow{12}{*}{S1} & \multirow{3}{*}{Task 1} & 5 & 4.7 & 0 & 90.7 & 92.7 & 33.8 & 91.6 & 94.4 & 93.8 \\
 &  & 50 & 76.1 & 0 & 67.9 & 92.7 & 13. & 88.8 & 93.4 & 94. \\
 &  & 100 & 96.1 & 0 & 53.3 & 92.2 & 3.4 & 85.8 & 92.8 & 94. \\ \cline{2-11} 
 & \multirow{3}{*}{Task   2} & 5 & 1. & 0 & 95.1 & 95.4 & 62.8 & 94.8 & 96.2 & 96.2 \\
 &  & 50 & 22.9 & 0 & 87.7 & 95.1 & 24.6 & 86.8 & 96. & 96.4 \\
 &  & 100 & 50.3 & 0.4 & 77.4 & 94.3 & 10.6 & 82.2 & 96. & 96.4 \\ \cline{2-11} 
 & \multirow{3}{*}{Task   3} & 5 & 10.7 & 0 & 93.7 & 94.2 & 11.8 & 90.8 & 98.6 & 95.2 \\
 &  & 50 & 83.1 & 3.1 & 63.3 & 91.7 & 7.2 & 79. & 97.4 & 95. \\
 &  & 100 & 97.5 & 13.5 & 46.4 & 88.9 & 2.6 & 69.2 & 97.6 & 94.6 \\ \cline{2-11} 
 & \multirow{3}{*}{Task   4} & 5 & 4.2 & 0 & 84.2 & 92.6 & 55. & 90.2 & 86.6 & 93.6 \\
 &  & 50 & 37.5 & 0 & 74.5 & 91.3 & 14.6 & 86.6 & 87.2 & 93.4 \\
 &  & 100 & 67.7 & 0.7 & 66.2 & 89.9 & 11.4 & 82.2 & 85.8 & 93.2 \\ \hline
\multirow{12}{*}{S2} & \multirow{3}{*}{Task 1} & 5 & 0 & 0 & 92. & 92.7 & 92. & 91.8 & 92.8 & 93.6 \\
 &  & 50 & 74.5 & 0 & 69.8 & 92.7 & 9.8 & 90. & 92.8 & 94.2 \\
 &  & 100 & 14.4 & 0.1 & 48.7 & 92.2 & 4.8 & 82. & 86.8 & 93.8 \\ \cline{2-11} 
 & \multirow{3}{*}{Task   2} & 5 & 1.6 & 0 & 94.8 & 95.4 & 38.6 & 94.6 & 96.2 & 96.2 \\
 &  & 50 & 22.4 & 0 & 87.8 & 95.1 & 20.8 & 89.6 & 96. & 96.2 \\
 &  & 100 & 44.5 & 0 & 79.2 & 94.6 & 15.8 & 87.2 & 95.8 & 96.4 \\ \cline{2-11} 
 & \multirow{3}{*}{Task   3} & 5 & 12.1 & 0 & 93.9 & 94.3 & 28. & 92.4 & 98.4 & 95. \\
 &  & 50 & 86. & 1. & 60.8 & 92.7 & 3.4 & 84. & 97.6 & 95. \\
 &  & 100 & 97.5 & 18.8 & 43.7 & 88.6 & 3.4 & 60.2 & 97.6 & 94.4 \\ \cline{2-11} 
 & \multirow{3}{*}{Task   4} & 5 & 3. & 0 & 84.5 & 92.6 & 48.8 & 91.2 & 86.6 & 93.4 \\
 &  & 50 & 32.8 & 0 & 76.1 & 91.1 & 24. & 86.8 & 86.8 & 93.4 \\
 &  & 100 & 63.8 & 2. & 66.5 & 89.6 & 17.4 & 80.6 & 86.8 & 93.2 \\ \hline
\multirow{12}{*}{S3} & \multirow{3}{*}{Task 1} & 5 & 0 & 0 & 92. & 92.7 & 92. & 91.8 & 92. & 93.6 \\
 &  & 50 & 77.7 & 0 & 68.4 & 92.6 & 12. & 90.2 & 92.8 & 93.8 \\
 &  & 100 & 96.5 & 0 & 56.4 & 92.3 & 3. & 90.2 & 89.4 & 93.4 \\ \cline{2-11} 
 & \multirow{3}{*}{Task   2} & 5 & 1.8 & 0 & 94.9 & 95.4 & 39. & 94.4 & 96.2 & 96. \\
 &  & 50 & 24.6 & 0 & 86.3 & 95.1 & 22.6 & 90. & 96.4 & 96.4 \\
 &  & 100 & 52.3 & 0 & 77.5 & 94.6 & 12. & 87.8 & 96. & 95.8 \\ \cline{2-11} 
 & \multirow{3}{*}{Task   3} & 5 & 10.5 & 0 & 94.4 & 94.3 & 32.8 & 92. & 98.4 & 95. \\
 &  & 50 & 81.9 & 2. & 65.8 & 91.9 & 4.2 & 78.4 & 98. & 95.2 \\
 &  & 100 & 97.1 & 13.9 & 49.4 & 89.3 & 3.8 & 62.2 & 97.4 & 94.8 \\ \cline{2-11} 
 & \multirow{3}{*}{Task   4} & 5 & 2.4 & 0 & 84.5 & 92.6 & 52. & 91.8 & 86.8 & 93.2 \\
 &  & 50 & 32. & 0 & 76. & 91.3 & 28. & 86. & 87. & 93.2 \\
 &  & 100 & 61.3 & 2. & 68.5 & 90. & 13.6 & 77.6 & 87. & 93.4 \\ \hline
\end{tabular}
}
\caption{Radiation tolerance of different memory configurations in sensitive areas.}
\label{tab:exp}
\end{table*}

\end{document}